\documentclass[epsfig,useAMS,usenatbib]{mn2e}
\usepackage{epsfig,graphics}
\newcommand{\be}{\begin{equation}}
\newcommand{\ee}{\end{equation}}
\title[Distant field BHB stars III]
{Distant Field BHB Stars III: Identification of a probable outer halo
  stream at Galactocentric distance r$ = 70\,$kpc}
\author[L. Clewley et al.]
  {L. Clewley$^{1}$\thanks{E-mail: clewley@astro.ox.ac.uk}, S. J. Warren$^{2}$, P. C. Hewett$^{3}$,
John. E. Norris$^{4}$, M. I. Wilkinson$^{3}$,
\newauthor N. W. Evans$^{3}$\\
$^{1}$Department of Physics, Denys Wilkinson Bldg., University of
Oxford, Keble Road, Oxford, OX1 3RH \\
$^{2}$Blackett Laboratory, Imperial College of Science Technology
and Medicine, Prince Consort Road, London SW7 2BW \\
$^{3}$Institute of Astronomy, Madingley Road, Cambridge CB3 0HA\\
$^{4}$Research School of Astronomy \& Astrophysics, The Australian National
University, Mount Stromlo Observatory, Cotter Road, \\ Weston, ACT 2611,
Australia}

\date{Released 2002 Xxxxx XX}

\pagerange{\pageref{firstpage}--\pageref{lastpage}} \pubyear{2002}

\def\LaTeX{L\kern-.36em\raise.3ex\hbox{a}\kern-.15em
    T\kern-.1667em\lower.7ex\hbox{E}\kern-.125emX}

\begin{document}
\label{firstpage}
\maketitle
\begin{abstract}
We present VLT-FORS1 spectra of a sample of 34 faint 20.0 $< g^{*} <$
21.1 A--type stars selected from the Sloan Digital Sky Survey Early
Data Release, with the goal of measuring the velocity dispersion of
blue horizontal branch (BHB) stars in the remote Galactic halo,
$R\sim80\,$kpc.  We show that colour selection with $1.08<
u^{*}-g^{*}<1.40$ and $-0.2< g^{*}-r^{*}<-0.04$ minimises
contamination of the sample by less luminous blue stragglers.  In
classifying the stars we confine our attention to the 20 stars with
spectra of signal-to-noise ratio $>15$\AA$^{-1}$.  Classification
produces a sample of eight BHB stars at distances $65-102\,$kpc from
the Sun (mean $80\,$kpc), which represents the most distant sample of
Galactic stars with measured radial velocities.  The dispersion of the
measured radial component of the velocity with respect to the centre
of the Galaxy is $58\pm15$km\,s$^{-1}$.  This value is anomalously low
in comparison with measured values for stars at smaller distances, as
well as for satellites at similar distances.  Seeking an explanation
for the low measured velocity dispersion, further analysis reveals
that six of the eight remote BHB stars are plausibly associated with a
single orbit.  Three previously known outer halo carbon stars also
appear to belong to this stream.  The velocity dispersion of all nine
stars relative to the orbit is only $15\pm4$km\,s$^{-1}$.  Further
observations along the orbit are required to trace the full extent of
this structure on the sky.

\end{abstract}


\section{Introduction} 

The existence of a dark massive halo appears to be a generic feature
of many galaxies, but the total masses, sizes, and the formation
history of galactic halos are poorly understood.  This is mostly
because we do not have large enough samples of dynamical tracers at
sufficiently large radii.  The formation and extent of such mass
distributions are of great importance in understanding the nature of
the dark matter and its role in galaxy formation and evolution.  For
instance, the quantification of the dark matter content of the Galaxy
would allow us to construct a picture of the assembly of the various
baryonic components, through comparison with simulations.  The
baryonic components can, in turn, provide information about the
evolution of the halo.

There is compelling evidence that at least part of the stellar halo
has been built up via the accretion of smaller satellite galaxies.
Numerous searches have been made for streams of material responsible
for building up the Galaxy.  A striking example is the identification
of the Sagittarius dwarf galaxy (Ibata, Gilmore \& Irwin, 1994) and
its stellar stream (e.g.  Helmi \& White, 2001).  Recently, an
extensive stream of stars has been uncovered in the halo of the
Andromeda galaxy (M31), revealing that it too is cannibalising a small
companion (e.g. Lewis et al.  2004).  Such streams yield crucial
information on the accretion history and formation of galaxy halos.
Extended stellar streams have also been used to constrain the mass of
the Galactic halo (e.g.  Johnston et al.  1999) and in M31 (Ibata et
al.  2004).  A number of authors have noted the possible evidence for
streams in the distribution of the intrinsically rare, but very
luminous, carbon stars.  Sanduleak (1980) proposed the association of
a single carbon star with the Magellanic Stream and Totten \& Irwin
(2000) made the general observation that the non-uniform distribution
of carbon-stars in their extensive survey of the halo may indicate
that a number of the stars are associated with streams.

We have previously argued (Clewley et al., 2002, hereafter Paper I)
that blue horizontal branch (BHB) stars are an ideal population for
exploring the outer reaches of the Halo.  Like carbon stars they are
luminous standard candles but are also far more numerous.

BHB stars are A--type giants.  A--type stars in the Galactic halo are
easily identified, as they lie blueward of the main--sequence turnoff
(e.g.  Yanny et al., 2000, hereafter Y2000).  Unfortunately assembling
clean samples of remote r $> 60\,$kpc~\footnote{In this paper we use
the coordinate $r$ to denote Galactocentric distances and the
coordinate $R$ to denote heliocentric distances.}  BHB stars is made
difficult by the existence of a contaminating population of
high--surface--gravity A--type stars, the blue stragglers, that are
between one and three mag.  fainter.  Previous analyses required high
signal-to-noise ratio ($S/N$) spectroscopy to reliably separate these
populations (e.g.  Kinman, Suntzeff, and Kraft, 1994), making
identification of BHB stars in the distant halo unfeasible.

This paper is the third in a series.  In Paper I we developed two
classification methods that enabled us to overcome the difficulties in
cleanly separating BHB stars from blue stragglers, and outlined an
observational programme to survey the halo for BHB stars.  In the
second paper (Clewley et al., 2004, hereafter Paper II), we presented
photometry and spectroscopy of faint $16.0<B<19.5$ candidate BHB stars
in two northern high Galactic latitude fields and four southern
fields.  This work resulted in a sample of 60 BHB stars at distances
$11<R<52\,$kpc (mean $28\,$kpc), with measured radial velocities.
Here we apply the methods of Papers I and II to survey for halo BHB
stars at much greater distances, $65<R<115\,$kpc.  The new survey uses
Sloan Digital Sky Survey (SDSS) photometry to isolate a sample of
faint halo A--type stars.  Reliable classifications are derived from
medium resolution spectroscopy using FORS1 at the VLT.

The candidate BHB stars observed at the VLT were selected using
$u^*g^*r^*$ photometry from the SDSS Early Data Release (EDR) data set
(Stoughton et al., 2002).  The EDR photometry was preliminary, and has
since been revised.  Because we need accurate Johnson-Kron-Cousins
$B,V$ magnitudes for the classification, we have used the more recent
SDSS Data Release 2 (DR2) photometry (Abazajian, 2004) of the
EDR--selected candidates for this purpose.  In Section 2 of this paper
we describe the selection of the BHB candidates from the EDR data set,
and provide our prescription for transforming the DR2 $g,r$ magnitudes
of A--type stars to $B,V$ magnitudes.  Section 3 provides a summary of
the VLT spectroscopic observations, and the data reduction and line
measurement procedures followed.  In Section 4 we use the methods of
Papers I and II to classify these stars into categories BHB and blue
straggler, and provide a summary table of distances and radial
velocities of the eight stars classified as BHB.  We compute the
velocity dispersions of the two populations and compare them with
previous work.  In Section 5 we discuss the kinematics of the BHB
stars.  We perform an orbital analysis of the sample and suggest that
most of them are plausibly associated with a single orbit.

\begin{figure}
\centering{
\scalebox{0.3}{
\includegraphics*[-50,140][700,700]{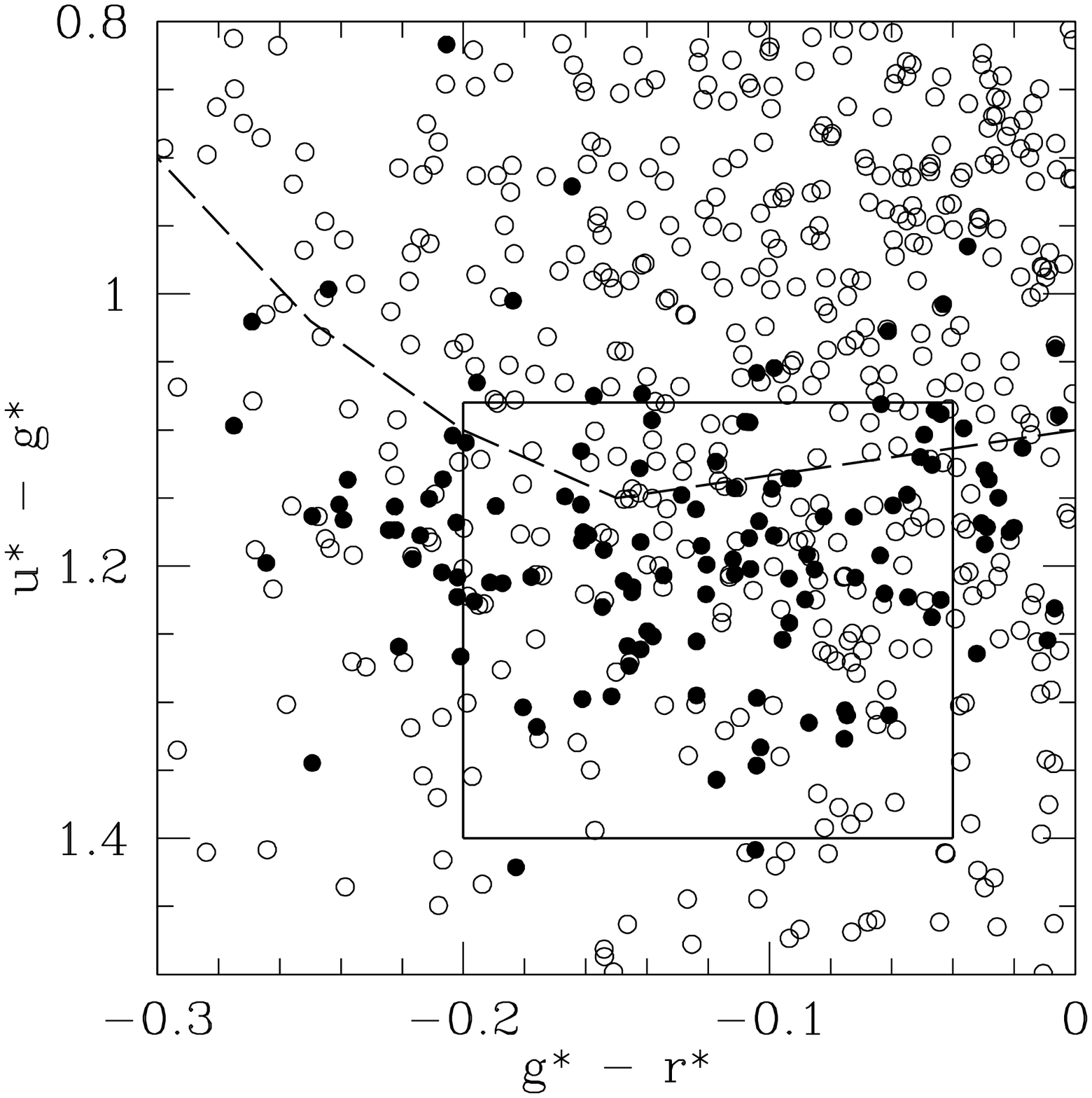}}
\scalebox{0.3}{
\includegraphics*[-50,140][700,700]{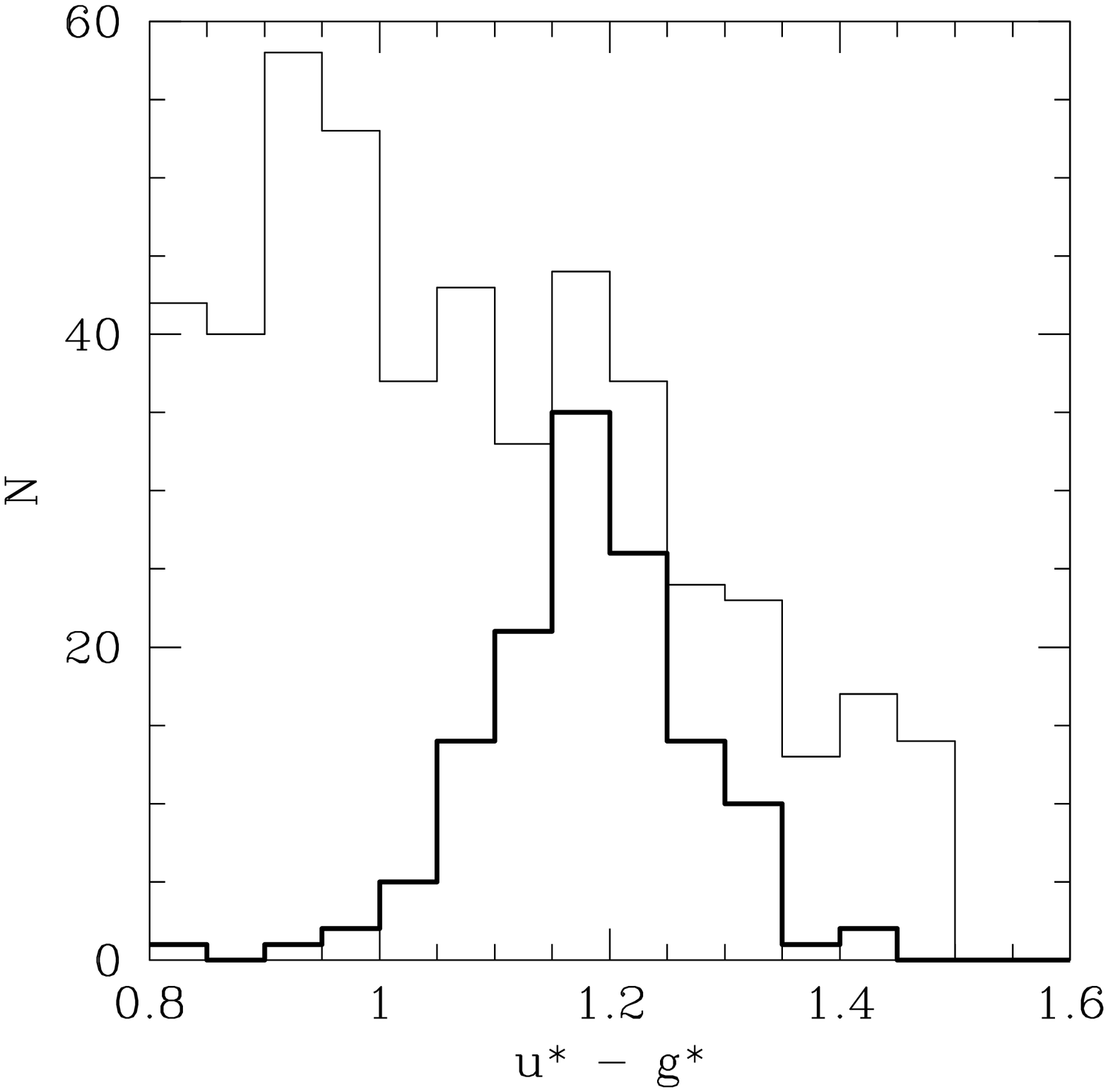}}
}

\caption{{\em Upper:}  Two colour plot of stellar objects in SDSS stripe 10, in
the RA range $200^\circ<\alpha<230^\circ$, the region including the
Sagittarius tidal stream.  Filled circles mark objects in the apparent
magnitude range $18.8<g^*<19.2$, expected to be predominantly BHB
stars.  Open circles mark objects in the apparent magnitude range
$20.5<g^*<21.5$, expected to be predominantly blue stragglers.  The
box marks our adopted colour selection.  The dashed line marks the
division between high and low gravity A stars employed by Y2000.  {\em
Lower:} Histograms of the $u^*-g^*$ colours of the stars in the upper
plot.  The thick line corresponds to the filled circles, and the thin
line to the open circles.  Typical colour errors for these two samples
are $\sigma(u^*-g^*)=0.05$ and 0.20, respectively.}

\label{select_ugr}
\end{figure}
\section{Selecting the BHB candidates}

\subsection{Colour selection}

We selected candidate BHB stars using the SDSS EDR point spread
function (PSF) $u^{*}g^{*}r^{*}$ magnitudes of stellar objects.  The
SDSS photometry has evolved between EDR and DR2 for a variety of
reasons: i) the reference photometric system is now that of the SDSS
2.5m telescope itself, rather than the photometric monitoring
telescope, ii) the absolute calibration of the standards has improved,
and the calibration of the survey data relative to the standards has
improved, iii) instrumental systematics (especially scattered light)
are now better understood.  Therefore, we have taken advantage of the
improved photometry in DR2 for the subsequent analysis, in particular
in classifying the objects and estimating their distances.  To
distinguish between photometric systems, the EDR system is designated
by asterisks, the system of the photometric monitoring telescope is
designated by primes, and the DR2 magnitudes are unadorned, which is
the SDSS convention.  Bearing in mind that all the stars are in the
remote halo, all the SDSS magnitudes discussed in this paper have been
corrected for Galactic extinction, using the map of Schlegel,
Finkbeiner \& Davis (1998).

We limited ourselves to the northern equatorial stripe in the EDR data
set, which covers $145^{\circ}<\alpha<236^{\circ}$,
$-1.25^{\circ}<\delta<+1.25^{\circ}$ (J2000).  This is SDSS stripe 10,
observed in runs 752 and 756 (Stoughton et al., 2002).  In selecting
candidate distant BHB stars from the EDR, we were guided by the
results of Y2000, who studied the spatial distribution of a sample of
A--type stars selected from the EDR using the (reddening--corrected)
colour selection box $-0.3<g^*-r^*<0.0$, $0.8<u^*-g^*<1.5$.  In
plotting apparent magnitude against $\alpha$ for these A--type stars,
Y2000 discovered that the distribution is not smooth, but includes
striking over--dense regions occurring in bands.  These have
subsequently been identified as tidal debris from the Sagittarius
dwarf galaxy.  Furthermore the bands occur in pairs, coincident on the
sky, but separated by $\sim2$mag.  The brighter objects are the more
luminous BHB stars, and the fainter objects are blue stragglers at the
same spatial location in the halo.

Our aim was to use the EDR data set to select a sample of candidate
faint BHB stars, with minimal contamination by blue stragglers, in
order to make the most efficient use of the spectroscopic time awarded
(for classification and velocity measurement).  To reach large
distances, we chose the magnitude range $20.0<g^*<21.1$, corresponding
to $\sim65<{\rm R}<\sim115\,$kpc, if the objects are BHB stars.
Fortunately, as demonstrated by Lenz et al.  (1998) using synthetic
photometry, the $u^*g^*r^*$ colours of A--type stars show some
dependence on surface gravity.  Given the accuracy of the photometry
at the distances of interest, $\sim0.1$mag., the $u^*g^*r^*$ colours
cannot provide {\em reliable} separation of the two populations.
Nevertheless Y2000 demonstrated that a colour cut in the $u^*-g^*$
{\it verus} $g^*-r^*$ plane is effective in enhancing the contrast of
the individual bands of tidal debris i.e.  can substantially reduce
the contamination of one population by the other.

Another way of looking at this is illustrated in Fig.  1.  The upper
diagram plots the colours of all EDR stars within the colour selection
box of Y2000, for the limited range
$\mathrm{200^\circ<\alpha<230^\circ}$, which is the region of the EDR
containing the strongest tidal debris bands.  Stars in the brighter
band $18.8<g^*<19.2$, which should be predominantly BHB stars, are
marked with solid symbols, while stars in the fainter band
$20.5<g^*<21.5$, which should be predominantly blue stragglers, are
marked with open symbols\footnote{The larger magnitude interval
selected is because blue stragglers have a larger spread in luminosity
than BHB stars (\S3.2).}.  For reference the dashed line shows the
dividing line used by Y2000.  The brighter band, typical colour error
$\sigma(u^*-g^*)=0.05$, is mostly confined to a narrow range in
$u^*-g^*$, which evidently defines the colour domain of the BHB stars.
The fainter stars, open symbols, are concentrated towards the top of
the plot, but are spread over a larger colour range.  However, much of
the spread is accounted for by the larger colour errors
$\sigma(u^*-g^*)=0.2$, as can be seen by reference to the histogram in
the lower plot.  It is evident that the mean $u^*-g^*$ colour of blue
stragglers is substantially bluer than for BHB stars.

The spectroscopic classification criteria (detailed in \S4) work best
near $(B-V)_0=0.1$, which corresponds to $g^{*}-r^{*}=-0.125$, using
the colour transformation provided by Fukugita et al.  (1996).  On
this basis we adopted the colour cuts shown by the box in Fig.  1,
defined by $1.08<u^*-g^*<1.40$, $-0.2<g^*-r^*<-0.04$, and limited
candidate selection to objects with colour error
$\sigma(g^*-r^*)<0.07$.  While these colour cuts should be nearly
optimal in terms of the fraction of candidates that are BHB stars, we
would still expect substantial contamination by blue stragglers, on
account of the large $u^*-g^*$ colour errors at the faint magnitudes
of the sample, $20.0<g^*<21.1$.

The distribution in $\alpha$ and $g^*$ of all the stars satisfying
these criteria is shown in Fig.  2.  The higher density of points at
$\alpha>200^\circ$ is due to blue stragglers in the Sagittarius tidal
stream.  Our classification methods produce samples of BHB stars that
are contaminated by blue stragglers at the level of $<10\%$, for
samples of A--type stars with a typical mix of the two populations
(Paper I).  The contamination of our BHB sample would be substantially
greater if we attempted to classify stars in this region, so we
confined our sample to $\alpha<200^\circ$.  The clump visible at
$\alpha=153^\circ$, $g^*\sim20.2$, is the horizontal branch of the
Sextans dwarf spheroidal \footnote{Note that the discussion in Y2000
of these objects mistakenly cites a paper concerned with a different
galaxy, Sextans A.}, centre $\alpha=153.3^\circ$,
$\delta=-1.61^\circ$, J2000 (Irwin and Hatzidimitriou, 1995).  In
order to avoid stars in Sextans we confined our selection to
$\alpha>160^\circ$.  The final selection includes 54 objects, of which
we observed the 35 listed in Table 1.  One of these, no.  32, proved
to be a quasar.  Looking ahead, of the remaining 34 candidates we are
able to classify 20.  These classifications are indicated on Fig.  2,
with solid circles representing the eight stars classified BHB, and
open circles representing the 12 stars classified blue straggler.  For
the remaining 14 candidates the classifications are uncertain, mostly
due to insufficient $S/N$.

In Table 1 we provide details of the 35 objects observed.  Column 1 is our
running number, and column 2 lists the coordinates.  Successive columns provide
the dereddened DR2 $g$ magnitude, and the dereddened $u-g$ and $g-r$ colours.
The last column provides the dereddened $B-V$ colour, calculated using the
transformation derived in \S2.2.  There is good agreement between the EDR and
DR2 photometry in the mean, but with noticeable scatter.  For example, looking
at the difference $(g^*-r^*)-(g-r)$ for our targets, the mean is 0.00 mag., and
the standard deviation is 0.03 mag.

\begin{figure}
\centering{
\scalebox{0.3}{
\includegraphics*[-50,140][700,700]{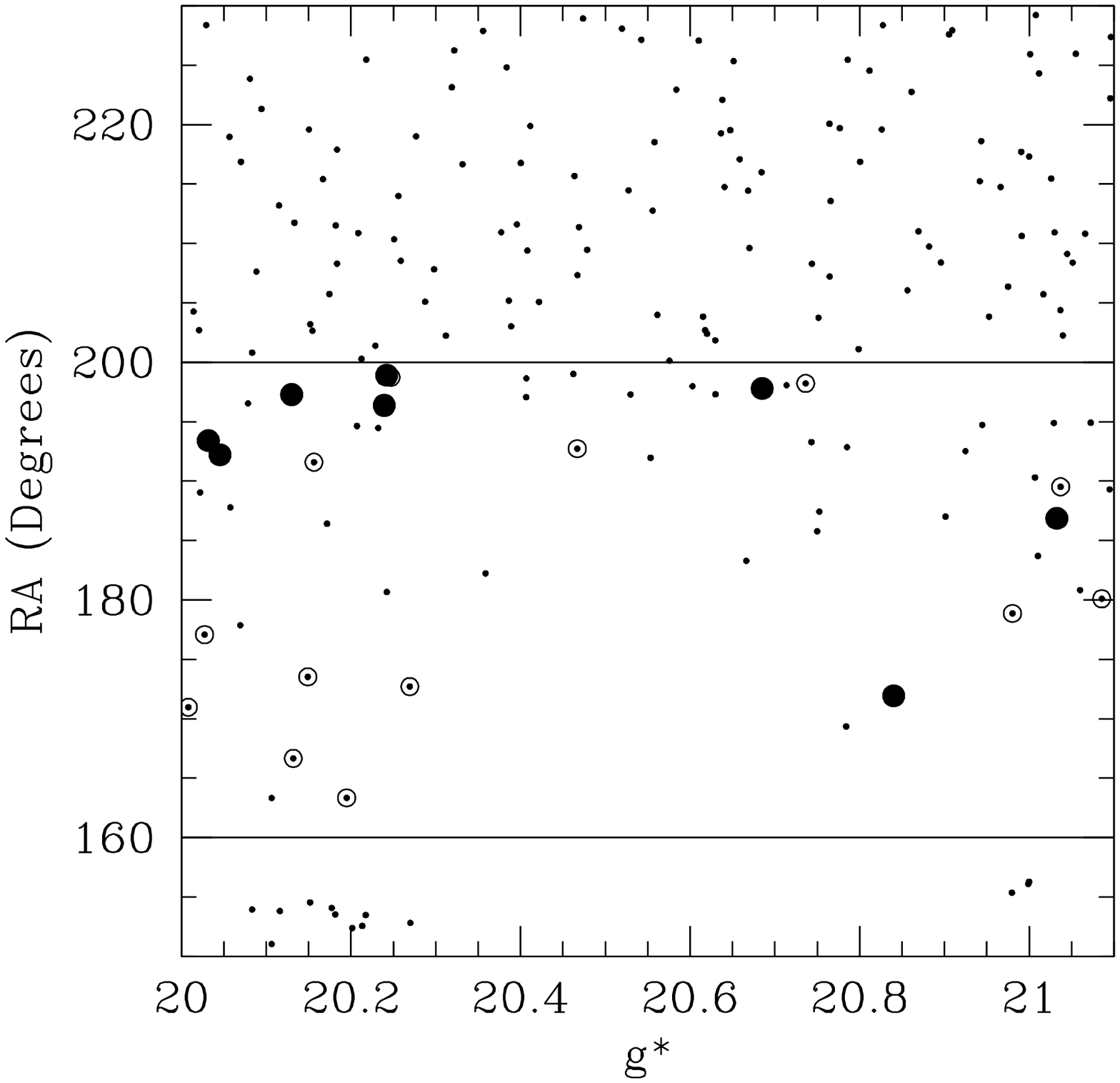}}
}

\caption{A plot of $\alpha$ against $g^*$ for all the BHB star candidates in EDR
stripe 10, satisfying our selection criteria (detailed in the text).
Selection was confined to the range $160^\circ<\alpha<200^\circ$.
There are 54 stars satisfying the selection criteria, of which the 35
listed in Table 1 were observed and 20 were classifiable.  The eight
stars classified as BHB stars are marked by filled circles, and the 12
stars classified as blue stragglers are marked by open circles.}

\label{select_ra_dec}
\end{figure}

\begin{table*}
\begin{center}
\begin{center}
\centering
\begin{tabular}{clcccc}
\hline
\noalign{\smallskip}   
No. & Identification (J2000) & $g$ & ($u-g$)$_0$ & ($g-r$)$_0$ & ($B-V$)$_0$ \\
\hline  
 01  & J$105319.31+004842.79$   &   20.09   $\pm$   0.02   &   1.08   $\pm$   0.09   &  -0.08   $\pm$   0.03   &   0.15   $\pm$   0.03   \\
 02  & J$105322.38-004449.73$   &   20.19   $\pm$   0.03   &   0.81   $\pm$   0.16   &  -0.10   $\pm$   0.05   &   0.14   $\pm$   0.04   \\
 03  & J$110639.10-004720.06$   &   20.16   $\pm$   0.03   &   1.11   $\pm$   0.15   &  -0.13   $\pm$   0.05   &   0.11   $\pm$   0.04   \\
 04  & J$112352.91-003719.83$   &   20.01   $\pm$   0.02   &   1.33   $\pm$   0.11   &  -0.16   $\pm$   0.03   &   0.09   $\pm$   0.03   \\
 05  & J$112744.35+001508.47$   &   20.85   $\pm$   0.03   &   1.29   $\pm$   0.18   &  -0.14   $\pm$   0.05   &   0.11   $\pm$   0.04   \\
 06  & J$113050.42-005147.11$   &   20.27   $\pm$   0.04   &   1.03   $\pm$   0.10   &  -0.14   $\pm$   0.05   &   0.11   $\pm$   0.04   \\
 07  & J$113407.58-004735.65$   &   20.21   $\pm$   0.04   &   1.28   $\pm$   0.19   &  -0.06   $\pm$   0.06   &   0.18   $\pm$   0.04   \\
 08  & J$114818.87-003921.03$   &   20.06   $\pm$   0.05   &   1.10   $\pm$   0.15   &  -0.08   $\pm$   0.07   &   0.16   $\pm$   0.05   \\
 09  & J$115525.99-003601.66$   &   20.96   $\pm$   0.03   &   1.27   $\pm$   0.24   &  -0.08   $\pm$   0.06   &   0.16   $\pm$   0.04   \\
 10  & J$120024.18+011026.81$   &   21.06   $\pm$   0.04   &   1.28   $\pm$   0.24   &  -0.13   $\pm$   0.07   &   0.12   $\pm$   0.05   \\
 11  & J$120855.70+010929.20$   &   20.35   $\pm$   0.03   &   1.18   $\pm$   0.12   &  -0.14   $\pm$   0.04   &   0.10   $\pm$   0.04   \\
 12  & J$121447.45+004001.27$   &   21.00   $\pm$   0.04   &   1.15   $\pm$   0.17   &  -0.10   $\pm$   0.06   &   0.14   $\pm$   0.05   \\
 13  & J$122305.31-011443.14$   &   20.75   $\pm$   0.04   &   1.22   $\pm$   0.21   &  -0.10   $\pm$   0.07   &   0.14   $\pm$   0.06   \\
 14  & J$122726.77+004641.03$   &   21.01   $\pm$   0.03   &   1.16   $\pm$   0.18   &  -0.09   $\pm$   0.06   &   0.15   $\pm$   0.05   \\
 15  & J$122802.26-010353.91$   &   20.93   $\pm$   0.05   &   1.10   $\pm$   0.22   &  -0.11   $\pm$   0.08   &   0.13   $\pm$   0.06   \\
 16  & J$123805.94+001941.07$   &   21.04   $\pm$   0.04   &   1.13   $\pm$   0.22   &  -0.06   $\pm$   0.07   &   0.17   $\pm$   0.05   \\
 17  & J$124112.31-010447.12$   &   21.03   $\pm$   0.04   &   1.06   $\pm$   0.22   &  -0.06   $\pm$   0.07   &   0.17   $\pm$   0.06   \\
 18  & J$124620.96-002802.04$   &   20.13   $\pm$   0.02   &   1.26   $\pm$   0.12   &  -0.20   $\pm$   0.04   &   0.06   $\pm$   0.03   \\
 19  & J$124851.75+003045.08$   &   20.05   $\pm$   0.02   &   1.21   $\pm$   0.10   &  -0.07   $\pm$   0.04   &   0.17   $\pm$   0.03   \\
 20  & J$125004.94+003422.85$   &   20.91   $\pm$   0.04   &   1.14   $\pm$   0.22   &  -0.10   $\pm$   0.06   &   0.14   $\pm$   0.05   \\
 21  & J$125054.28-000759.84$   &   20.44   $\pm$   0.03   &   1.12   $\pm$   0.11   &  -0.17   $\pm$   0.04   &   0.08   $\pm$   0.04   \\
 22  & J$125123.59+000345.91$   &   20.78   $\pm$   0.03   &   1.01   $\pm$   0.15   &  -0.11   $\pm$   0.05   &   0.13   $\pm$   0.04   \\
 23  & J$125308.36-000554.68$   &   20.75   $\pm$   0.03   &   1.20   $\pm$   0.17   &  -0.08   $\pm$   0.05   &   0.15   $\pm$   0.04   \\
 24  & J$125336.47-002414.48$   &   20.01   $\pm$   0.03   &   1.16   $\pm$   0.09   &  -0.14   $\pm$   0.04   &   0.11   $\pm$   0.03   \\
 25  & J$125934.17+002058.64$   &   21.07   $\pm$   0.04   &   1.00   $\pm$   0.22   &  -0.07   $\pm$   0.07   &   0.17   $\pm$   0.06   \\
 26  & J$125938.97+000748.09$   &   21.07   $\pm$   0.04   &   1.56   $\pm$   0.32   &  -0.10   $\pm$   0.07   &   0.14   $\pm$   0.05   \\
 27  & J$130528.14+004855.90$   &   20.23   $\pm$   0.04   &   1.14   $\pm$   0.11   &  -0.06   $\pm$   0.05   &   0.17   $\pm$   0.04   \\
 28  & J$130907.25+005731.22$   &   20.11   $\pm$   0.03   &   1.16   $\pm$   0.11   &  -0.10   $\pm$   0.05   &   0.14   $\pm$   0.04   \\
 29  & J$131109.31+000950.36$   &   20.67   $\pm$   0.03   &   1.28   $\pm$   0.20   &  -0.18   $\pm$   0.05   &   0.07   $\pm$   0.04   \\
 30  & J$131155.98+003914.93$   &   20.61   $\pm$   0.03   &   0.84   $\pm$   0.13   &  -0.11   $\pm$   0.05   &   0.13   $\pm$   0.04   \\
 31  & J$131252.50+000821.47$   &   20.72   $\pm$   0.03   &   1.05   $\pm$   0.16   &  -0.13   $\pm$   0.05   &   0.11   $\pm$   0.04   \\
 32  & J$131435.78+010329.82$Q  &   20.55   $\pm$   0.03   &   0.98   $\pm$   0.16   &   0.01   $\pm$   0.06   &   0.22   $\pm$   0.04   \\
 33  & J$131458.12-004706.39$   &   20.30   $\pm$   0.03   &   1.11   $\pm$   0.14   &  -0.08   $\pm$   0.04   &   0.15   $\pm$   0.04   \\
 34  & J$131538.01-010853.30$   &   20.26   $\pm$   0.03   &   1.25   $\pm$   0.13   &  -0.08   $\pm$   0.04   &   0.16   $\pm$   0.04   \\
 35  & J$131602.90+010150.60$   &   20.50   $\pm$   0.03   &   1.24   $\pm$   0.17   &  -0.10   $\pm$   0.05   &   0.14   $\pm$   0.04   \\
\noalign{\smallskip}    						       
\hline 
\end{tabular}
\end{center}
\caption{Photometric data for the 35 BHB candidates from the SDSS DR2.
One candidate that was subsequently found to be a quasar has `Q' appended to
the SDSS identifier.
\label{col_mags}}
\end{center}
\end{table*}

\subsection{Transforming from $g-r$ to $B-V$}\label{sec:trans}

We use two methods to classify stars into categories BHB star and blue
straggler.  One method makes use of $(B-V)_0$ colours.  Therefore we
need to convert the extinction corrected $g-r$ colours to $B-V$.
Fukugita et al (1996) quote the transformation \be
B-V=(g^\prime-r^\prime+0.23)/1.05, \ee computed from synthetic
photometry.  Smith et al.  (2002) quote the transformation \be
B-V=(g^\prime-r^\prime+0.19)/0.98, \ee based on observations of
standard stars.  These two relations are plotted in Fig.  3, as the
dashed and dotted lines respectively.  The two linear relations differ
somewhat even over the narrow range of colours of interest in this
paper, $-0.2<g-r<-0.04$, by some 0.03 mag.  Whereas one would
naturally prefer the empirical relation over the synthetic relation,
there is some indication that the actual transformation is non--linear
in the region of the A stars, as the four stars measured by Smith et
al.  (2002) in the colour range of interest, lie on average 0.04 mag.
above the linear relation (which is a fit over a wide colour range).
Further evidence that the relation is non--linear comes from our own
photometry, which is less precise, but has many more stars.  Plotted
in Fig.  1 are the colours of the 60 stars in the colour range $-0.05
< (B-V)_0 < 0.40$, with $(B-V)_0$ measured by ourselves (Paper II),
which also have SDSS DR2 $g-r$ colours.  Only the errors on $(B-V)_0$
are plotted, as these dominate.  Our data points are systematically
high relative to the linear relation of Smith et al.  (2002).  An
additional source of uncertainty is the fact that the above linear
relations were derived for the $u^\prime g^\prime r^\prime i^\prime
z^\prime$ system of the photometric monitoring telescope, slightly
different from the $ugriz$ system of the SDSS 2.5m telescope itself,
used for DR2.

Accurate $(B-V)_0$ photometry is required for the classification of
the stars using the {\em $D_{0.15}$--Colour} method (\S4).  A
systematic error in the $(B-V)_0$ colour as large as 0.05 mag.  could
result in many of the classifications being in error.  Therefore we
re-investigated the colour transformation.  We computed synthetic
colours using the methods detailed in Hewett et al.  (in prep.).  We
used model stars with [Fe/H]$= -1$ and log~$g$=3.5 (an appropriate
surface gravity midway between BHB stars and blue stragglers), from
Kurucz (1993).  We fit a cubic polynomial to the relation between the
$(B-V)_0$ and $(g-r)_0$ synthetic colours, for the colour range of
Fig.  3.  Finally, bearing in mind the uncertainty in the absolute
calibration of the SDSS magnitudes onto the AB system (Fukugita et
al., 1996), we allowed the zero point of the relation to be a free
parameter, established by shifting the derived curve vertically to
give the best fit to the data of Fig.  3. The curve provides a better
fit than the two linear relations plotted.  Furthermore the average offset
of the four stars measured by Smith et al.  (2002) in the colour range
of interest reduces to 0.005 mag.  We have therefore adopted this
relation.  The transformation is given by \be B-V = 0.764(g-r) -
0.170(g-r)^2 + 0.715(g-r)^3 + 0.218,
\ee and is plotted in Fig.  3 as the bold solid line.  We stress that
this relation is specifically for A--type stars, and is not expected
to be reliable for other types of star.  The adjustment to the zero
point was very small, only 0.014 mag.  The agreement is encouragingly
good and gives considerable confidence in the many elements going into
this comparison \---\ the SDSS DR2 photometry, the absolute
calibration of Vega, and of the SDSS standard stars, our own
photometry, the measurement of the different passband response
functions, and the synthetic stellar spectra.  The computed $(B-V)_0$
colours of our targets are listed in the final column of Table
\ref{col_mags}.

In summary, we have presented evidence that, in the colour range of
interest, A stars lie systematically off the simple linear colour
transformation measured by Smith et al.  (2002), by about 0.05 mag.
and we have derived a cubic relation that provides an improved fit.

\begin{figure}
\rotatebox{0}{ \centering{ \scalebox{0.4}{
\includegraphics*{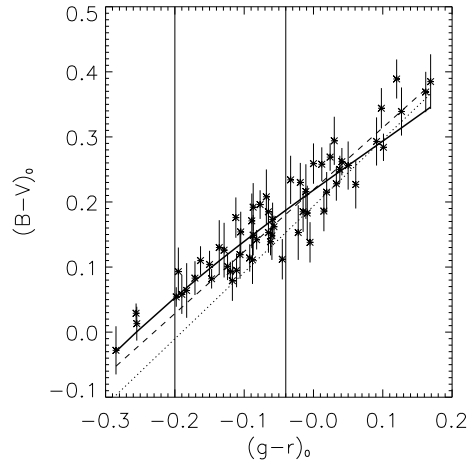} }}}

\caption{Colour transformation between dereddened SDSS DR2 $(g-r)_0$ colour and
$(B-V)_0$.  The data plot the photometry of the 60 stars from Paper II
with SDSS photometry.  The dashed line is the transformation derived
by Fukugita et al.  (1996) from synthetic photometry, and the dotted
line is the transformation measured by Smith et al.  (2002).  The
solid line is the cubic relation derived here from synthetic colours
of Kurucz (1993) model stars.  The vertical lines mark the
$(g^*-r^*)_0$ colour selection limits.
\label{rg_trans}}
\end{figure}

\section{Spectroscopic observations, analysis, and results}

\subsection{Observations}

We used the VLT FORS1 instrument, in service mode, over two periods,
from 2003/03/25 to 2003/04/09, and from 2004/01/30 to 2004/03/20, to
obtain medium resolution optical spectra of the 35 BHB candidates.
The instrument is equipped with a $2048^2$ Tek CCD, with a projected
scale of 0.2\arcsec pixel$^{-1}$.  We used the 600B grating, giving a
dispersion of 1.2 {\AA} pixel$^{-1}$.  With the 0.7\arcsec slit, the
resolution achieved, measured from arc lines, was about 4{\AA}, which
is sufficient for the line--fitting procedure.  The spectral coverage
was 3400--5700 {\AA}, which includes the relevant lines H$\delta$,
H$\gamma$, and Ca II K $\lambda3933$\AA.  Three BHB radial velocity
standard stars in the globular cluster M5 were observed twice each.
Table \ref{m5_gc} summarises relevant information on the standards.
Columns (1) to (5) list the identification, RA and Dec., $V$
magnitude, $(B-V)_{0}$ colour, and the heliocentric radial velocity,
V$_{\odot}$.  The information in successive columns (6) to (11) in
Table 3 contains averages of $H\delta$ and $H\gamma$ line
measurements.  Columns (6) to (8) in Table \ref{m5_gc} list,
respectively, the parameters $D_{0.15}$, $b$, and $c$ (explained in
\S4).  The errors on the parameters $b$ and $c$ are provided in
columns (9) to (11) in the form of $A$ and $B$, the semi-major and
semi-minor axes of the error ellipse in the $b-c$ plane, and $\theta$
the orientation of the semi-major axis, measured anti-clockwise from
the $b$-axis.  Here the error corresponds to the $68\%$ confidence
interval for each axis in isolation (see Paper I for further details).

The requested integration times of between 765 and 1980 seconds, for stars in
the magnitude range $20.0<g^*<21.1$, were estimated using the FORS1
exposure-time calculator, on the basis of the requested seeing, transparency,
and lunar phase, in order to achieve the minimum continuum $S/N$ ratio of 15
${\mathrm\AA}^{-1}$ required to classify the stars (Paper I).  All targets were
observed near culmination, with a mean airmass of $1.20\pm0.1$.  In the event,
observations of 12 of the 34 targets failed to achieve the required $S/N$, and
therefore these targets cannot be reliably classified.  The failures were
primarily in the cases where the seeing was poor.  In retrospect, for these
service observations we should have included a safety margin in the requested
integration times.  All image frames were automatically bias and flat--field
corrected by the FORS pipeline, and we then followed standard procedures for sky
subtraction, spectral extraction, and wavelength calibration.  An error
spectrum, used for the line profile fits, was computed from Poisson
considerations.

For the wavelength calibration, HgCdHe arc observations were made during the
day, and were used to derive the dispersion solution.  To account for any
flexure of the instrument, the zero point of the dispersion solution was
established for each spectrum using the [OI] night-sky line at 5577.34 {\AA}.
We found a {\em rms} drift of the zero point of $13\,$kms$^{-1}$ over the entire
data set.

\subsection{Analysis}

As stated earlier, one of the candidates proved to be a quasar.  In the
remainder of the paper we ignore this object, and refer only to the 34 stars.
The spectra were used to measure the shapes and widths of the H$\delta$ and
H$\gamma$ lines (for classification), the EW of the CaII K line (to measure the
metallicity), and the radial velocity of each star (for future dynamical
analysis).  For these measurements we followed the procedures set out in Papers
I and II exactly, and we refer the reader to those papers for full details.
Below we provide a brief explanation of how the Balmer lines were measured, and
how the CaII K line EW is used to determine the metallicity.  We then summarise
how the magnitude, colour, and metallicity are combined with the classification
to estimate a distance for each star.

{\em Balmer line profiles.}  After normalising each spectrum to the continuum,
we fit a S\'{e}rsic function, convolved with a Gaussian of FWHM the instrumental
resolution, to the H$\delta$ and H$\gamma$ lines.  Two parameters of the fit,
the scale width $b$, and the shape index $c$, are recorded.  One classification
procedure, the {\em Scale width--Shape} method, plots these two quantities
against each other.  A third quantity $D_{0.15}$, which is the line width at a
depth $15\%$ below the continuum, is derived from $b$ and $c$.  The second
classification method, the {\em $D_{0.15}-$colour} method, plots this quantity
against $(B-V)_0$.  Because $D_{0.15}$ is a function of $b$ and $c$, the two
classification methods are not completely independent.

{\em Metallicities from CaII K lines.}  The CaII K line is the strongest metal
line present in the wavelength range covered by the spectra, and the only useful
line in moderate resolution blue spectra for measuring metallicity.  Plotting
CaII K line EW against $(B-V)_0$, the metallicity is determined by interpolation
between lines of constant metallicity on this diagram (see Fig.  4).  The
uncertainty is established from the uncertainties of the two quantities plotted,
and an additional uncertainty of 0.3dex is added in quadrature.  This is the
systematic error, and was established by comparing metallicities derived by this
method using high $S/N$ data of comparable resolution, with accurate
metallicities determined from high--resolution spectra.  No attempt has been
made to remove the possible contribution of interstellar CaII K absorption from
the stellar K measurements.  For a remote halo star the typical CaII K EW is
0.11\AA/sin$b$ (Bowen, 1991), with a $95\%$ range of $(0.06-0.31)$\AA/sin$b$.
This range translates to $0.07-0.35$\AA\, for our fields.  Our 34 stars have
mean EW 1.7\AA\, and standard deviation 0.8\AA, with only two stars having EW
below 0.7\AA.  Therefore interstellar CaII K absorption is insignificant for the
majority of our targets, but could bias the measured metallicities high for the
small fraction of stars with the weakest lines.

{\em Distances.}  The absolute magnitude of a BHB star $M_V(BHB)$ depends on
both metallicity and colour (i.e.  temperature).  In Paper II we derived a
relation for the absolute magnitude of BHB stars by combining published
relations for the dependence of $M_V$ on metallicity and on $(B-V_0)$ colour,
with the measured absolute magnitude at fixed metallicity and colour, as
follows.  The slope of the relation between apparent magnitude and metallicity
for RR Lyrae stars in the Large Magellanic Cloud was measured by Clementini
(2003) from observations of some 100 stars.  We combined this with the
measurement of the absolute magnitude of RR Lyrae stars at fixed metallicity,
determined by Gould \& Popowski (1998) from Hipparcos statistical parallaxes, to
derive the linear relation for RR Lyrae stars:

\begin{equation} 
M_V(RR)=1.112+0.214[\mathrm{Fe/H}].
\label{ch4:abs_mag}
\end{equation} 

We then adopted a cubic expression determined by Preston et al. (1991),
for the $(B-V)_0$ colour dependence of the difference in absolute
magnitudes between BHB and RR Lyrae stars, to produce the final
expression for the absolute magnitude of BHB stars:
\begin{eqnarray}  
M_V(BHB) &=& 1.552+0.214[\mathrm{Fe/H}]-4.423(B-V)_0 \nonumber\\ & & +
17.74(B-V)^2_0-35.73(B-V)^3_0.
\label{ch4:abs_mag}
\end{eqnarray}

Distances and associated errors are then determined using the apparent
magnitudes $V_0$, and the corresponding photometric and metallicity errors.  To
compute $V$, we used the relation $V=g^\prime-0.53(g^{\prime}-r^{\prime})$
(Fukugita et al., 1996), here disregarding the subtle differences between the
different SDSS magnitudes ($g, g^{\prime}, g^*$, etc.).  The result produces
distance errors of $6-10\%$ for our confirmed BHB stars.  The exact form of
Equation \ref{ch4:abs_mag}, particularly the zero point, remains controversial.
Currently there are at least ten methods of determining the absolute magnitudes
of RR Lyrae stars.  We refer the interested reader to a recent review of this
subject by Cacciari \& Clementini (2003).\footnote{They find, by averaging over
all ten methods in their review,

\begin{equation}  
M_V(RR) = 0.93 \pm 0.12 + (0.23 \pm 0.04)[\mathrm{Fe/H}]
\end{equation}}

\begin{figure}
\centering{ \scalebox{0.3}{
\includegraphics*[-50,140][700,700]{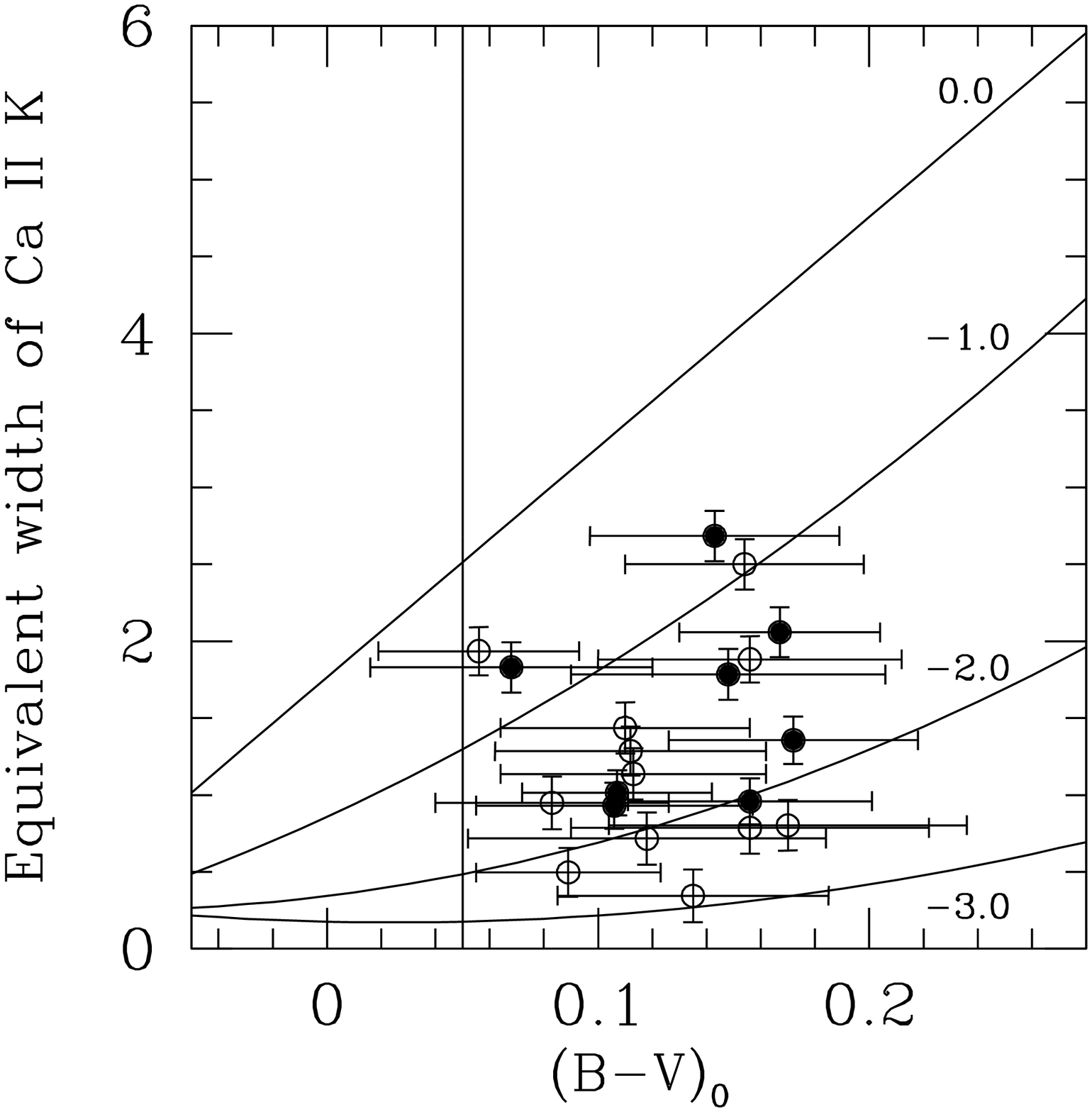}}}
\caption{CaII K line ($3933\,$\AA) EW(\AA) for the sample plotted against
$(B-V)_0$. The curves represent lines of constant metallicity for
[Fe/H] = -1.0, -2.0 and -3.0 taken from Wilhelm et al. (1999). The
straight line represents a best fit to stars in the Pleiades and Coma
clusters assumed to be of solar metallicity. The vertical line at
$(B-V)_0$ = 0.05 is the limit for which metallicities can be
determined. The 8 stars classified as BHB stars are marked by filled
circles, and the 12 stars classified blue straggler are marked by open
circles.}
\label{fig_cak}
\end{figure}

The absolute magnitudes of blue stragglers have been less well studied.
Since we will not use the blue stragglers in any dynamical analysis,
their distances are less interesting. In Paper II we adopted the
following relation derived by KSK from data for blue stragglers in
globular clusters published by Sarajedini (1993)
\be
M_V(BS)=1.32+4.05(B-V)_0-0.45[\mathrm{Fe/H}].
\ee

\begin{figure*}
\begin{minipage}{133mm}
\epsfig{figure=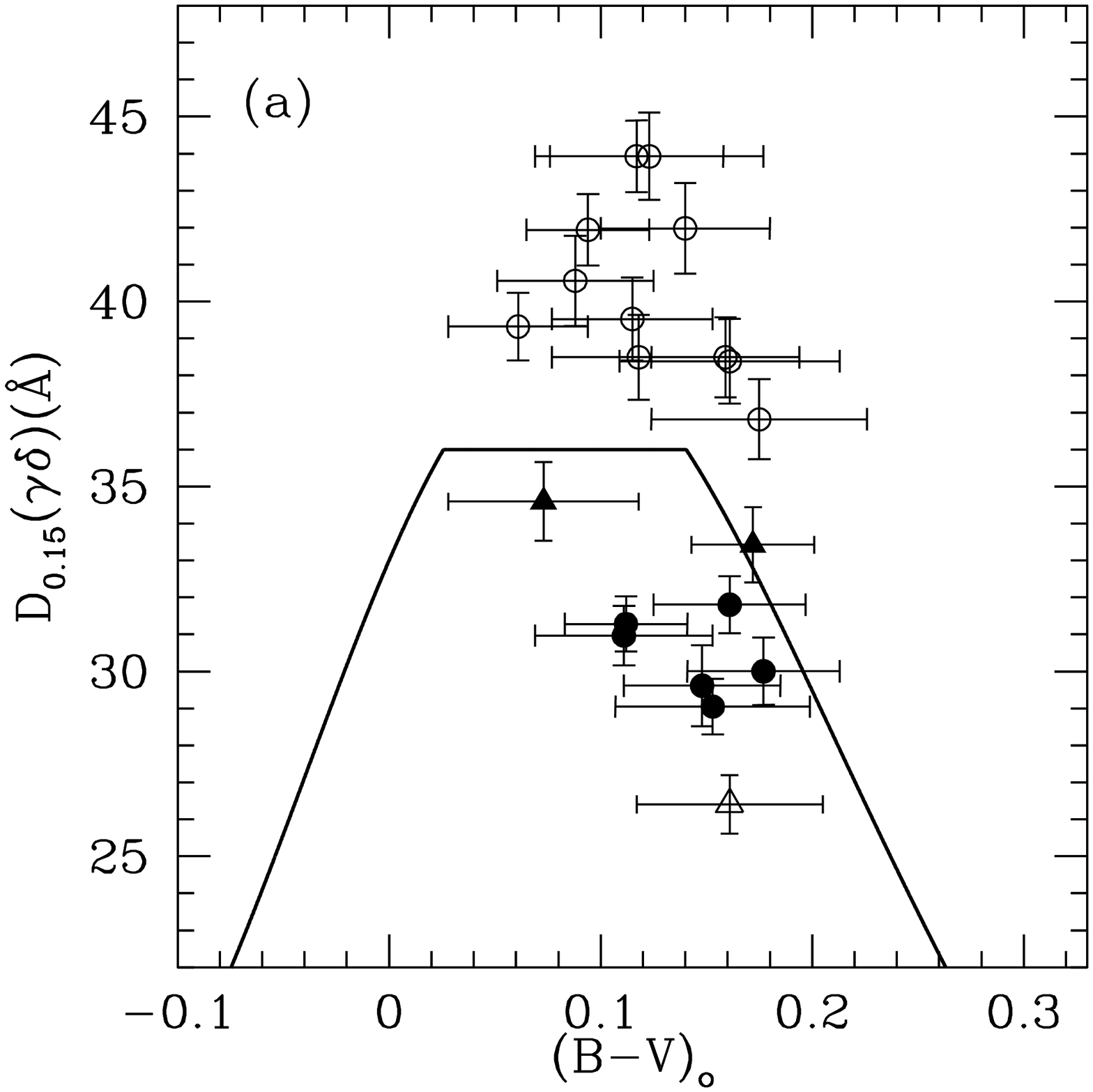,height=66mm,width=66mm}
\epsfig{figure=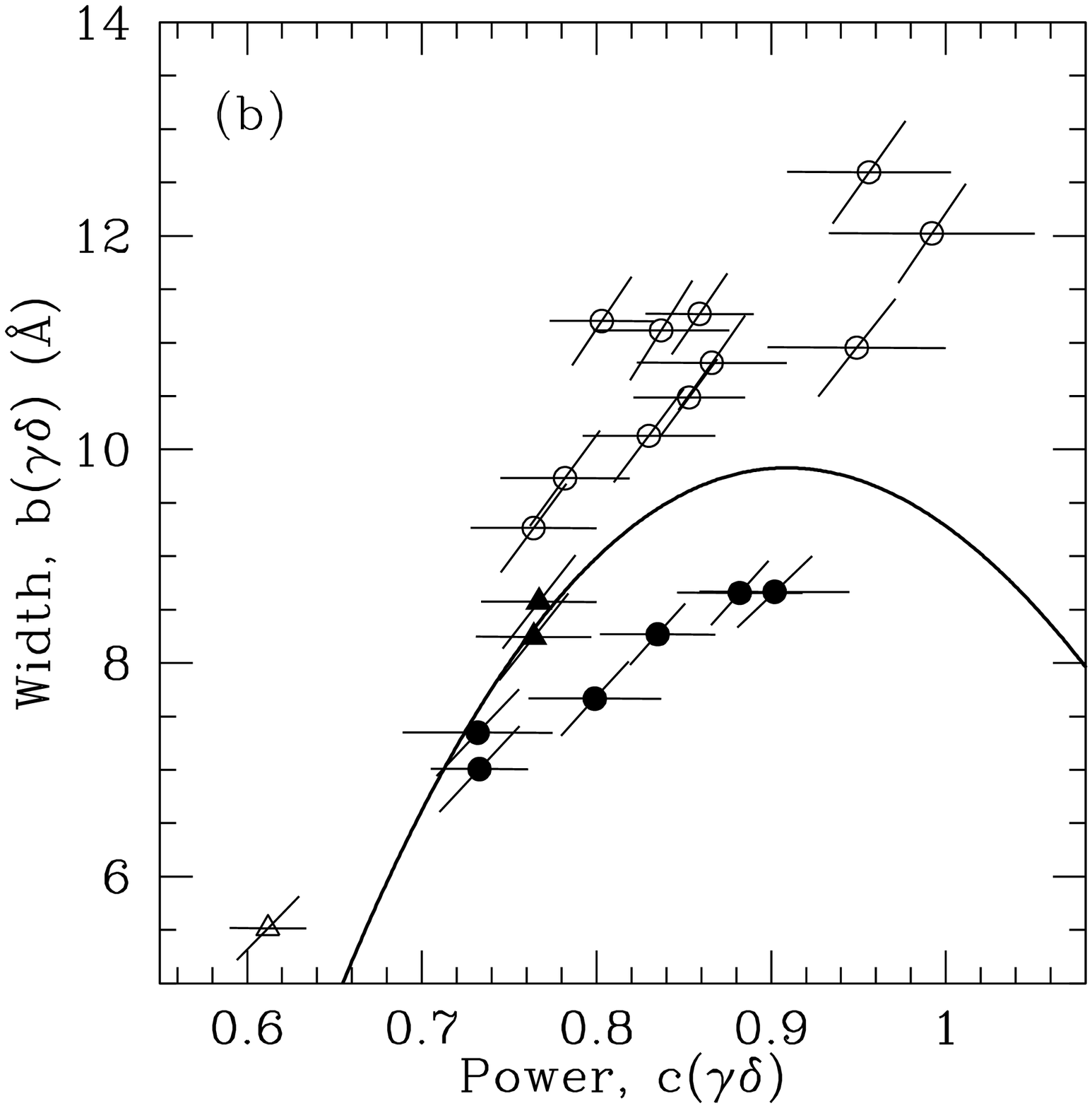,height=66mm,width=66mm}
\end{minipage}
\caption{Classification of the 20 survey stars, using the {\em
$D_{0.15}$--Colour} (a) and the {\em Scale width--Shape} (b)
classification methods. The solid curves are the classification
boundaries explained in the text. Filled circles are stars classified
BHB in both plots, i.e. the stars below the classification boundary in
each plot. Filled triangles are stars below the classification
boundary in only one plot but are nonetheless classified BHB, whereas
open triangles are classified A/BS. In plot (a) there are eight stars
below the boundary and 12 above it. In plot (b) there are seven stars
below the boundary and 13 above it. A total of eight stars are
classified BHB.
\label{sw_cw}}
\end{figure*}
\begin{table*}   
\begin{flushleft}
\begin{center}
\begin{tabular}{lcccccccccc}
\hline
\noalign{\smallskip}   
\multicolumn{1}{c}{ID} &   
\multicolumn{1}{c}{RA (J2000) Dec.} &   
\multicolumn{1}{c}{$V$} &     
\multicolumn{1}{c}{$(B-V)_0$} &   
\multicolumn{1}{c}{V$_{\odot}$} &   
\multicolumn{1}{c}{$D_{0.15}(\gamma\delta)$} & 
\multicolumn{1}{c}{$b(\gamma\delta)$} &       
\multicolumn{1}{c}{$c(\gamma\delta)$} &
\multicolumn{1}{c}{$A$} &
\multicolumn{1}{c}{$B$} & 
\multicolumn{1}{c}{$\theta$} \\   
\multicolumn{1}{c}{} &  
\multicolumn{1}{c}{} &
\multicolumn{1}{c}{} &
\multicolumn{1}{c}{} &   
\multicolumn{1}{c}{[km s$^{-1}$]} &      
\multicolumn{1}{c}{[\AA]} &      
\multicolumn{1}{c}{[\AA]} &    
\multicolumn{1}{c}{} &
\multicolumn{1}{c}{} &
\multicolumn{1}{c}{} &  
\multicolumn{1}{c}{} \\    
\noalign{\smallskip}   
\hline  
\noalign{\smallskip}   
M5-I-53  & J$151836.35+020744.6$ & 15.06 & 0.06 & 52.2 $\pm$ 1.4 & 29.47 $\pm$ 0.26 & 8.22 & 0.91 & 0.11 & 0.014 & 1.511  \\ 
         &                       &       &      &                & 29.38 $\pm$ 0.28 & 8.14 & 0.90 & 0.11 & 0.014 & 1.511  \\ \hline 
M5-II-78 & J$151826.93+020717.8$ & 14.95 & 0.12 & 42.2 $\pm$ 1.1 & 27.81 $\pm$ 0.28 & 7.58 & 0.87 & 0.11 & 0.015 & 1.509  \\ 
         &                       &       &      &                & 28.22 $\pm$ 0.28 & 7.65 & 0.87 & 0.12 & 0.014 & 1.511  \\ \hline
M5-IV-05 & J$151835.34+020227.9$ & 15.15 & 0.15 & 56.9 $\pm$ 1.2 & 29.51 $\pm$ 0.28 & 8.07 & 0.88 & 0.11 & 0.014 & 1.512  \\ 
         &                       &       &      &                & 29.66 $\pm$ 0.27 & 8.12 & 0.89 & 0.11 & 0.013 & 1.512  \\ \hline
\noalign{\smallskip}   
\end{tabular}
\caption{Spectroscopic measurements of three M5 globular cluster BHB
  stars. The names are from Arp (1955) and Arp (1962), the photometry
  is from Cudworth (1979), and the radial velocities are from Peterson
  (1983).\label{m5_gc}} 
\end{center}
\end{flushleft}
\end{table*}

\subsection{Results}

The results of these measurements for the 34 candidate BHB stars are
provided in Table \ref{spec_tabs}.  Column (1) gives the number of the
star, and columns (2) and (3) record the EW of the H$\gamma$ line, and
the spectrum continuum $S/N$ per \AA.  Our classification
methods were developed specifically for objects with strong Balmer
lines, defined by EW H$\gamma>13$\AA, and with continuum
$S/N>15$\AA$^{-1}$.  In all, only 20 of the 34 candidates meet both
criteria, and are therefore classifiable.  The majority fail because
of inadequate $S/N$.  The information in successive columns (4) to (9)
in Table 3 contain averages of $H\delta$ and $H\gamma$ line
measurements.  The quantities in columns (4) to (9) in Table 3, i.e.
$D_{0.15}$, $b$, and $c$, $A$, $B$ and $\theta$, are also provided for
the radial velocity standards in columns (6) to (11) of Table 2.
Comparing these quantities enables us to use the observations of the
standard stars as a further check of our classification methods.

In column (10) of Table \ref{spec_tabs} is listed the EW of the CaII K line.
This is plotted against $(B-V)_0$ in Fig.  \ref{fig_cak}.  The measured
metallicity for each star, and the error (including random and systematic
contributions) is provided in column (11).  The large errors are a consequence
of the comparatively large errors in the $g^*-r^*$ colours.  The mean measured
metallicity of the stars plotted is $-1.4$ with dispersion 0.6, similar to the
mean value measured for our sample of brighter A--type stars (Paper II).  There
are no significant outliers, but this is not a strong statement, given the large
errors.

The radial velocity, corrected to the heliocentric frame, is provided in column
(12), and the estimated distance, based on the classification from the following
section, is provided in column (13).

\begin{table*}
\begin{center}
\begin{tiny}
\begin{tabular}{lrrcrccccrcrrl}
\hline
\noalign{\smallskip}   
\multicolumn{1}{c}{No.} &   
\multicolumn{1}{c}{EW($\gamma$)} &     
\multicolumn{1}{c}{S/N} &   
\multicolumn{1}{c}{$D_{0.15}(\gamma\delta)$} &   
\multicolumn{1}{c}{$b(\gamma\delta)$} &    
\multicolumn{1}{c}{$c(\gamma\delta)$} &
\multicolumn{1}{c}{$A$} &  
\multicolumn{1}{c}{$B$} &  
\multicolumn{1}{c}{$\theta$} &    
\multicolumn{1}{c}{EW(CaIIK)} &   
\multicolumn{1}{c}{[Fe/H]} & 
\multicolumn{1}{c}{V$_{\odot}$} &     
\multicolumn{1}{c}{$R$} &   
\multicolumn{1}{c}{prob./class.} \\
\multicolumn{1}{c}{} &  
\multicolumn{1}{c}{[\AA]} &
\multicolumn{1}{c}{[\AA]$^{-1}$} &  
\multicolumn{1}{c}{[\AA]} &
\multicolumn{1}{c}{[\AA]} &
\multicolumn{1}{c}{} &  
\multicolumn{1}{c}{} &  
\multicolumn{1}{c}{} &  
\multicolumn{1}{c}{} &
\multicolumn{1}{c}{[\AA]} &   
\multicolumn{1}{c}{} &   
\multicolumn{1}{c}{[km s$^{-1}$]} &  
\multicolumn{1}{c}{[kpc]} &  
\multicolumn{1}{c}{} \\
\noalign{\smallskip}   
\hline  
\noalign{\smallskip}
(1)&  (2) &  (3) &       (4)      &  (5)  &  (6) &  (7) &  (8)  &  (9)  &    (10)       &    (11)          &   (12)        &   (13)     &   (14)         \\
01 & 19.2 & 10.3 & 35.78$\pm$1.69 &  9.24 & 0.77 & 0.68 & 0.068 & 1.521 & 0.92$\pm$0.19 &  -2.09$\pm$0.41  & 86.6$\pm$19.0 & 73.4$\pm$5.6 & (0.39)\,A/BS? \\  
02 & 19.8 & 15.2 & 41.98$\pm$1.23 & 11.12 & 0.84 & 0.47 & 0.039 & 1.532 & 0.34$\pm$0.17 &  -2.87$\pm$0.37  & 393.8$\pm$12.4 & 26.6$\pm$7.9 & (0.00)\,A/BS \\  
03 & 17.7 & 15.9 & 38.50$\pm$1.15 & 10.95 & 0.95 & 0.46 & 0.051 & 1.523 & 1.14$\pm$0.17 &  -1.64$\pm$0.41  & 260.9$\pm$12.7 & 36.3$\pm$5.1 & (0.09)\,A/BS \\  
04 & 20.0 & 19.2 & 41.94$\pm$0.97 & 11.27 & 0.86 & 0.38 & 0.031 & 1.529 & 0.50$\pm$0.16 &  -2.32$\pm$0.38  & 172.0$\pm$10.9 & 30.2$\pm$5.6 & (0.00)\,A/BS \\  
05 & 16.1 & 20.4 & 30.97$\pm$0.81 &  8.67 & 0.90 & 0.33 & 0.043 & 1.507 & 0.93$\pm$0.15 &  -1.78$\pm$0.45  & 172.8$\pm$16.4 & 98.5$\pm$7.9 & (0.97)\,BHB \\   
06 & 18.8 & 15.7 & 39.53$\pm$1.12 & 10.81 & 0.87 & 0.44 & 0.043 & 1.528 & 1.44$\pm$0.17 &  -1.37$\pm$0.43  & 354.3$\pm$13.8 & 39.2$\pm$5.5 & (0.04)\,A/BS \\  
07 & 19.1 & 14.5 & 38.75$\pm$1.13 &  9.44 & 0.71 & 0.45 & 0.033 & 1.527 & 1.35$\pm$0.17 &  -1.83$\pm$0.49  & -35.5$\pm$09.8 & 30.3$\pm$6.8 & (0.03)\,A/BS? \\ 
08 & 18.8 & 17.9 & 38.38$\pm$1.14 &  9.73 & 0.78 & 0.45 & 0.037 & 1.526 & 0.79$\pm$0.17 &  -2.29$\pm$0.51  & 56.9$\pm$16.5 & 27.0$\pm$7.1 & (0.08)\,A/BS \\   
09 & 14.2 & 21.2 & 26.40$\pm$0.79 &  5.52 & 0.61 & 0.30 & 0.022 & 1.511 & 1.88$\pm$0.15 &  -1.35$\pm$0.48  & 45.0$\pm$09.4 & 49.8$\pm$7.6 & (0.33)\,A/BS \\   
10 & 20.4 & 15.0 & 43.93$\pm$1.18 & 12.60 & 0.96 & 0.48 & 0.047 & 1.527 & 0.72$\pm$0.17 &  -2.11$\pm$0.55  & 64.5$\pm$15.4 & 47.8$\pm$9.9 & (0.00)\,A/BS \\   
11 & 19.0 & 12.6 & 39.75$\pm$1.49 &  9.68 & 0.74 & 0.59 & 0.040 & 1.528 & 1.59$\pm$0.18 &  -1.20$\pm$0.42  & 133.1$\pm$47.4 & 43.0$\pm$5.0 & (0.05)\,A/BS? \\ 
12 & 11.6 & 17.5 & 24.59$\pm$0.89 &  5.71 & 0.70 & 0.34 & 0.034 & 1.507 & 2.55$\pm$0.15 &  -0.81$\pm$0.54  & 96.7$\pm$15.2 & 98.1$\pm$6.6 & unclassifiable \\ 
13 & 16.8 & 12.3 & 35.49$\pm$1.38 &  9.14 & 0.77 & 0.55 & 0.050 & 1.520 & 2.18$\pm$0.18 &  -1.08$\pm$0.57  & 160.2$\pm$17.6 & 49.0$\pm$6.9 & (0.28)\,A/BS? \\ 
14 & 13.7 & 21.4 & 29.05$\pm$0.75 &  7.01 & 0.73 & 0.40 & 0.040 & 1.514 & 2.19$\pm$0.11 &  -1.14$\pm$0.48  & -21.2$\pm$10.5 & 102.3$\pm$5.1 & (0.75)\,BHB \\  
15 & 16.8 &  9.0 & 32.74$\pm$1.87 &  7.95 & 0.74 & 0.77 & 0.059 & 1.518 & 2.86$\pm$0.20 &  -0.50$\pm$0.71  & 19.4$\pm$24.5 & 91.4$\pm$6.5 & (0.52)\,BHB? \\   
16 & 17.4 & 15.8 & 36.82$\pm$1.08 &  9.26 & 0.76 & 0.42 & 0.036 & 1.526 & 0.80$\pm$0.16 &  -2.36$\pm$0.49  & 2.7$\pm$10.7 & 40.3$\pm$10.9 & (0.15)\,A/BS \\   
17 &  9.9 & 14.7 & 21.03$\pm$1.17 &  4.18 & 0.57 & 0.40 & 0.042 & 1.492 & 2.98$\pm$0.11 &  -0.80$\pm$0.61  & 97.4$\pm$16.9 & 101.5$\pm$7.3 & unclassifiable \\
18 & 19.2 & 19.9 & 39.32$\pm$0.91 & 10.49 & 0.85 & 0.36 & 0.032 & 1.526 & 1.94$\pm$0.16 &  -0.49$\pm$0.46  & 58.3$\pm$10.0 & 49.4$\pm$3.3 & (0.01)\,A/BS \\   
19 & 15.0 & 16.8 & 33.42$\pm$1.02 &  8.24 & 0.76 & 0.41 & 0.033 & 1.522 & 2.06$\pm$0.16 &  -1.31$\pm$0.41  & 38.3$\pm$12.2 & 68.7$\pm$4.8 & (0.65)\,BHB \\    
20 & 18.4 & 13.1 & 35.98$\pm$1.41 &  8.63 & 0.72 & 0.56 & 0.041 & 1.525 & 1.28$\pm$0.18 &  -1.68$\pm$0.47  & 120.2$\pm$13.4 & 46.2$\pm$8.5 & (0.24)\,A/BS? \\ 
21 & 19.2 & 15.7 & 40.56$\pm$1.21 & 12.02 & 0.99 & 0.47 & 0.059 & 1.529 & 0.95$\pm$0.17 &  -1.63$\pm$0.38  & -95.6$\pm$20.0 & 42.1$\pm$5.9 & (0.01)\,A/BS \\  
22 & 20.2 & 13.2 & 38.64$\pm$1.57 &  8.84 & 0.67 & 0.60 & 0.037 & 1.530 & 3.13$\pm$0.19 &  -0.38$\pm$0.55  & 74.2$\pm$22.5 & 58.5$\pm$5.1 & (0.06)\,A/BS? \\  
23 & 16.1 & 13.9 & 34.92$\pm$1.33 &  7.96 & 0.68 & 0.54 & 0.035 & 1.525 & 1.73$\pm$0.18 &  -1.44$\pm$0.45  & -31.5$\pm$16.6 & 44.2$\pm$7.2 & (0.29)\,A/BS? \\ 
24 & 13.7 & 21.9 & 31.28$\pm$0.75 &  8.27 & 0.84 & 0.29 & 0.033 & 1.516 & 1.01$\pm$0.15 &  -1.71$\pm$0.36  & 18.0$\pm$09.7 & 66.8$\pm$5.0 & (0.96)\,BHB \\    
25 & 12.2 & 16.1 & 24.90$\pm$1.08 &  5.55 & 0.66 & 0.41 & 0.037 & 1.508 & 2.79$\pm$0.16 &  -0.87$\pm$0.59  & 32.6$\pm$16.8 & 104.3$\pm$7.3 & unclassifiable  \\
26 & 13.2 & 12.3 & 31.07$\pm$1.40 &  7.27 & 0.71 & 0.52 & 0.043 & 1.521 & 2.07$\pm$0.18 &  -1.12$\pm$0.53  & 78.3$\pm$16.9 & 105.1$\pm$8.0 & (0.48)\,A/BS? \\ 
27 & 13.8 & 18.9 & 30.00$\pm$0.91 &  7.67 & 0.80 & 0.35 & 0.038 & 1.515 & 1.36$\pm$0.15 &  -1.81$\pm$0.43  & 90.3$\pm$08.9 & 78.5$\pm$6.5 & (0.94)\,BHB \\    
28 & 15.2 & 16.0 & 29.61$\pm$1.09 &  7.35 & 0.73 & 0.41 & 0.043 & 1.512 & 2.68$\pm$0.16 &  -0.74$\pm$0.49  & 88.8$\pm$11.5 & 64.9$\pm$4.0 & (0.68)\,BHB \\    
29 & 15.9 & 16.5 & 34.59$\pm$1.06 &  8.57 & 0.77 & 0.44 & 0.033 & 1.523 & 1.83$\pm$0.16 &  -0.69$\pm$0.54  & 92.3$\pm$12.7 & 76.8$\pm$4.8 & (0.51)\,BHB \\    
30 & 10.5 & 11.2 & 25.17$\pm$1.34 &  6.22 & 0.76 & 0.53 & 0.059 & 1.509 & 3.66$\pm$0.18 &  -0.04$\pm$0.54  & 47.4$\pm$21.3 & 77.0$\pm$4.3 & unclassifiable  \\
31 & 20.1 & 19.0 & 43.93$\pm$0.97 & 11.20 & 0.80 & 0.42 & 0.030 & 1.530 & 1.28$\pm$0.16 &  -1.51$\pm$0.44  & 41.2$\pm$11.3 & 47.0$\pm$6.7 & (0.00)\,A/BS \\   
33 & 18.2 & 16.3 & 38.49$\pm$1.09 & 10.13 & 0.83 & 0.44 & 0.038 & 1.525 & 2.50$\pm$0.16 &  -0.96$\pm$0.45  & 133.7$\pm$12.9 & 40.1$\pm$5.0 & (0.08)\,A/BS \\  
34 & 15.0 & 20.0 & 31.80$\pm$0.77 &  8.66 & 0.88 & 0.30 & 0.036 & 1.516 & 0.96$\pm$0.15 &  -2.05$\pm$0.46  & 119.2$\pm$12.3 & 80.0$\pm$6.6 & (0.98)\,BHB \\   
35 & 11.6 & 13.5 & 27.81$\pm$1.16 &  6.04 & 0.63 & 0.44 & 0.037 & 1.513 & 1.28$\pm$0.17 &  -1.66$\pm$0.44  & 104.5$\pm$12.6 & 84.5$\pm$6.6 & unclassifiable   \\
\noalign{\smallskip} 
\hline 
\end{tabular}
\caption{Spectroscopic data for the horizontal branch star
candidates. \label{spec_tabs}}
\end{tiny}
\end{center}
\end{table*}

\section{Classification and velocity dispersion}

\subsection{Classification}

As noted above, of the 34 candidates, only 20 meet the requirements on
spectroscopic $S/N$ and H$\gamma$ EW for reliable classification.  In the
following we restrict our discussion to the classification of these 20 objects.
Of the other 14 candidates, five stars have EW H$\gamma<13$\AA, and are
considered unclassifiable.  For the remaining nine candidates the $S/N$ of the
spectra is too low for the classification to be reliable.  We have nevertheless
followed the classification procedures for these objects, but for clarity have
omitted them from Figures 4 and 5.  The final classifications are flagged as
questionable.  We have followed the classification procedures of Paper II (which
are slightly different from those of Paper I) exactly, with the exception that
we weight the two classification methods unequally, as detailed below.

In Figure \ref{sw_cw} we plot the two diagnostic diagrams for the 20 classifiable
stars in the survey.  The two figures are explained as follows.  Figure
\ref{sw_cw}(a) shows the {\em $D_{0.15}-$colour} method.  The average values of
$D_{0.15}$ for H$\gamma$ and H$\delta$ against $(B-V)_0$ are plotted for the 20
candidates.  In Paper I we showed that reliable classification by this method
requires the uncertainty on $(B-V)_0$ to be less than 0.03 mag.  Unfortunately
this is untrue for most of the stars in our sample (Table 1).  For this reason
we give this method lower weight in the final classification.  Figure
\ref{sw_cw}(b) shows the {\em Scale width--Shape} method.  The line--profile
quantities $b$ and $c$, averaged for H$\gamma$ and H$\delta$ are
plotted.  The solid lines show the classification boundaries, from
Paper II, with high--surface gravity stars (i.e.  main--sequence A
stars or blue stragglers, hereafter A/BS) above the line, and
low--surface gravity stars (i.e.  BHB stars) below the line.  In both
plots stars classified BHB are plotted as solid symbols and stars
classified A/BS are plotted open.  As we discuss below, the three
triangles are stars that have ambiguous classifications, i.e stars
that are classified as BHB by one classification method and not the
other.

Inspection of Figure \ref{sw_cw} provides the following information.
Of the 20 candidates, eight are classified BHB by the {\em
$D_{0.15}-$colour} method.  The {\em Scale width--Shape} method
classifies seven stars as BHB.  There are six stars classified BHB by
both methods.  A total of nine stars are classified BHB by one or
other of the methods.  There is clearly close agreement between the
two classification methods, but there are three stars with ambiguous
classifications.  Before considering these further, we note that the
three radial velocity standards (Table 2), previously classified BHB
from high--resolution spectroscopy, are all unambiguously classified
BHB in both plots.

The uncertainties on each parameter define the 2D probability
distribution functions for any point.  By integrating these functions
below the classification boundary we can compute a probability
$P(BHB)$ that any star is BHB.  We can then average the probabilities
for the two classification methods, to improve the classification.  We
have computed these probabilities for each star, giving twice the
weight to the {\em Scale width--Shape} method when averaging (because,
as mentioned above, the {\em $D_{0.15}-$colour} method is affected by
the relatively large colour errors).  As in previous papers, stars
with $\bar{P}(BHB)>0.5$ are then classified BHB, and stars with
$\bar{P}(BHB)\le0.5$ are classified A/BS.  Based on the Monte Carlo
simulations of Paper I, we would expect the sample of BHB stars
defined in this way to be contaminated by A/BS stars at no more than
the $10\%$ level, which we consider satisfactory.  Col.  (14) of Table
3 provides the averaged probabilities, and corresponding
classifications, for the 20 classifiable stars.

Of the three stars with ambiguous classifications, 2 are classified
BHB.  The third star classified A/BS, star 9, is the object with the
smallest value of $D_{0.15}$, and the smallest value of $c$.  The
small value of $c$ indicates a colour substantially redder than
(although compatible with) the measured value of
$(B-V)_0=0.16\pm0.04$.

We also provide the classification probabilities for the nine stars
with inadequate spectroscopic $S/N$, but for these the classifications
are given as BHB?  or A/BS?  to indicate that they are not reliable.
Finally the five stars with EW H$\gamma<13$\AA\, are labelled
unclassifiable.

\begin{figure}
\rotatebox{0}{ 
\centering{ 
\scalebox{0.4}{
\includegraphics*{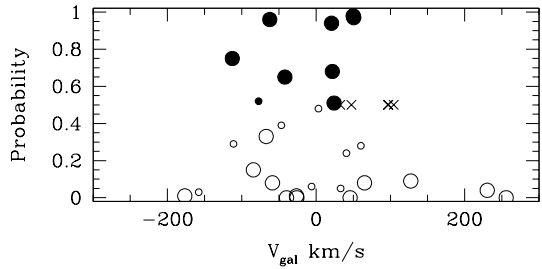}}}}

\caption{Classification probability against $V_{gal}$ for all the BHB
candidates.  Large circles represent the 20 stars with spectra of high {\em
S/N}.  The BHB stars which have $\bar{P}>0.5$ are shown by filled symbols and
A/BS stars are marked by open symbols.  Also plotted, as small symbols, are the
nine stars with unreliable classifications.  Finally the five unclassifiable
stars are marked by small crosses, at $\bar{P}=0.5$.
\label{prob_sn}}
\end{figure}

\subsection{Velocity dispersion}

Table \ref{data_sum} contains a summary of the kinematic properties of
the final sample of eight BHB stars.  Listed there are the Galactic
coordinates $l$ and $b$, and the Galactocentric radial velocity and
distance, V$_{gal}$ and $r$ respectively.  To convert the
heliocentric quantities to Galactocentric quantities, the heliocentric
radial velocities are first corrected for solar motion by assuming a
solar peculiar velocity of ($U,V,W$) = (-9,12,7), where $U$ is
directed outward from the Galactic Centre, $V$ is positive in the
direction of Galactic rotation at the position of the Sun, and $W$ is
positive toward the North Galactic Pole.  We have assumed a circular
speed of 220 km s$^{-1}$ at the Galactocentric radius of the Sun
($R_{\odot} = 8.0\,$kpc).  Table \ref{data_sum}, then, distills the
main observational result of the paper, a sample of distant BHB stars
with measured radial velocities.  The Table also includes three carbon
stars, designated by their coordinates, which are introduced in
Section 5.

We find, after quadratically subtracting the measurement errors in the
same manner as Norris \& Hawkins (1991), that the measured dispersion
of the radial component of the Galactocentric velocity dispersion for
our BHB sample is 58$\pm$15km s$^{-1}$ at a mean heliocentric distance
of $80\,$kpc.  In Table
\ref{ftest} we compare this value against the measured velocity dispersion of a
variety of samples.  The sample of remote BHB stars is referred to as Sample A,
and listed in the first line of Table \ref{ftest}.  The first comparison sample,
Sample B, comprises the 60 BHB stars $11<R<52\,$kpc, mean $R=28\,$kpc, from
Paper II, which is the largest sample at such distances.  Sirko et al.  (2004a)
have also isolated large samples of distant BHB stars using the SDSS.  They
split their sample into a bright ($g<18$) subsample, which is contaminated by
blue stragglers at the level of about 10\% (i.e.  similar to the work presented
here), and a faint subsample ($g>18$), which is contaminated at about 25\%.  If
we consider only their clean bright sample, here Sample C, then $\sigma =
99.4 \pm4.3\,$km\,s$^{-1}$ (Sirko et al.  2004b), at mean distance $16\,$kpc.
Sample D consists of the 12 stars in Table 3 classified BS, with measured
velocity dispersion $129\pm26\,$km\,s$^{-1}$, at mean distance $40\,$kpc.
Finally considering the Galactic satellites discussed in \S1, selecting the nine
satellites within the distance range of our remote BHB sample, i.e.  $65 < {\rm
R} < 102\,$kpc, Sample E, we measure a velocity dispersion $134 \pm 32\,$km
s$^{-1}$, at a mean distance $82\,$kpc.  The velocity dispersions, mean
distances, and sample sizes, of these four samples are entered in columns $2-4$
of Table \ref{ftest}.  In the final column we list the probability that the
measured value for the remote BHB stars could be drawn from the same population
as each of the four comparison samples, as measured by the F--test.  At better
than the $95\%$ significance level, our sample of remote BHB stars has smaller
velocity dispersion than BHB stars at much smaller radii ($16\,$kpc, Sample C),
BS stars of similar apparent magnitude at intermediate radii ($40\,$kpc, Sample
D), and satellites at the same radii ($82\,$kpc, Sample E).  Compared to the
sample of BHB stars at $28\,$kpc, the difference is not significant.  We
conclude that the velocity dispersion of the remote BHB stars is anomalously
low, and in the following section we seek an explanation.

\begin{table} 
\begin{flushleft}
\begin{center}
\begin{tabular}{lrrrr}
\hline
\noalign{\smallskip} 
\multicolumn{1}{c}{No.} & 
\multicolumn{1}{c}{$l$} &  
\multicolumn{1}{c}{$b$} & 
\multicolumn{1}{c}{$V_{gal}$} &
\multicolumn{1}{c}{$r$} \\
\multicolumn{1}{c}{} & 
\multicolumn{1}{c}{[$^{\circ}$]} &  
\multicolumn{1}{c}{[$^{\circ}$]} & 
\multicolumn{1}{c}{[km s$^{-1}$]} & 
\multicolumn{1}{c}{[kpc]} \\
\noalign{\smallskip} 
\hline 
\noalign{\smallskip} 
19 & 301.495 & 63.377 & -42.0 & 67.2 \\
24 & 304.106 & 62.463 & -62.3 & 65.1 \\
27 & 310.805 & 63.473 & 20.8 & 76.6 \\
28 & 312.874 & 63.489 & 21.9 & 62.9 \\
29 & 313.702 & 62.623 & 24.2 & 74.7 \\
34 & 315.536 & 61.130 & 49.9 & 77.6 \\
1225-0011 & 290.267 & 61.822 & -27.0 & 65.0 \\ 
1241+0237 & 298.280 & 65.159 & -32.0 & 66.0 \\
1249+0146 & 303.159 & 64.372 & -133.0 & 53.0 \\
\hline
05 & 262.690 & 56.439 & 50.7 & 99.4 \\
14 & 289.614 & 63.028 & -112.7 & 101.4 \\
\noalign{\smallskip} 
\hline 
\end{tabular}
\caption{Summary of kinematic information for the BHB and carbon stars
that are associated with the stream. The two BHB stars below the line
do not belong to the stream. \label{data_sum}}
\end{center}
\end{flushleft}
\end{table}

\begin{table*} 
\begin{flushleft}
\begin{center}
\begin{tabular}{clcccc}
\hline
\noalign{\smallskip} 
\multicolumn{1}{c}{Sample} &
\multicolumn{1}{c}{Population} & 
\multicolumn{1}{c}{Dispersion} &  
\multicolumn{1}{c}{$<R>$} & 
\multicolumn{1}{c}{Size} & 
\multicolumn{1}{c}{F--test} \\
\multicolumn{1}{c}{} & 
\multicolumn{1}{c}{} & 
\multicolumn{1}{c}{[km s$^{-1}$]} &  
\multicolumn{1}{c}{[kpc]} & 
\multicolumn{1}{c}{} & 
\multicolumn{1}{c}{[prob.]} \\
\noalign{\smallskip} 
\hline 
\noalign{\smallskip} 
A & BHB & $58\pm15$   & 80 & 8 & ... \\
B & BHB & $108\pm10$   & 28 & 60 & 0.09 \\
C & BHB & $99.4\pm4.3$ & 16 & 733 & 0.02  \\
D & BS  & $129\pm26$   & 40 & 12   & 0.02  \\
E & satellites & $134\pm32$ & 82 & 9 & 0.04\\ 
\noalign{\smallskip} 
\hline 
\end{tabular}
\caption{Comparison of measured velocity dispersion of a variety of
  populations in the Galactic halo, against the measured velocity
  dispersion of our new sample of eight remote BHB stars at mean distance
  ($<R>$) of $80\,$kpc, Sample A.
\label{ftest}}
\end{center}
\end{flushleft}
\end{table*}

\section{Discussion}

In considering the anomalously low velocity dispersion of the remote
BHB stars, we first check that the significantly different velocity
dispersion between the BHB and BS stars is robust.  In Fig.
\ref{prob_sn} we plot classification probability against $V_{gal}$.
In this plot, the large symbols represent the 20 stars with spectra of
high {\em S/N}.  BHB stars have $\bar{P}>0.5$, and, as usual, are
shown by filled symbols, with A/BS stars marked by open symbols.  Also
plotted, as small symbols, are the nine stars with unreliable
classifications.  Finally the five unclassifiable stars are marked by
small crosses, at $\bar{P}=0.5$.  This plot shows a clear difference
in the kinematics of stars at the bottom of the diagram (high
probability A/BS, large velocity spread), and at the top of the
diagram (high probability BHB, small velocity spread).  There is no
evidence for any BHB stars with large values of $|{V_{gal}}|$ that
have been missed, because they fall just outside the classification
boundary, or because they were unclassifiable because the spectra are
of low {\em S/N}.  Therefore, the difference in velocity dispersion
between the two populations is quite robust to the method of
classification.

Another concern we had was the possibility that the sample is
contaminated by misclassified blue stragglers in the Sagittarius
stream.  Fig.  2 plots $\alpha$ against $g^*$ of the initial list of
candidate BHB stars.  As we discussed earlier this colour--selected
sample of candidates is expected to be contaminated by blue
stragglers, because of the large photometric errors at these faint
magnitudes, and this is apparently confirmed by the high--density of
stars at $\alpha>200^\circ$, where presumably most of the stars are
blue stragglers.  It was therefore worrying that the eight BHB stars,
marked in the upper diagram by filled circles, mostly lie close to the
boundary $\alpha=200^\circ$ in this plot, whereas one might expect
them to be more uniformly scattered over the RA range.  If these stars
are misclassified blue stragglers, on the edge of the Sagittarius
stream, this would provide a natural explanation for the small
velocity dispersion.  However, if there are any Sagittarius blue
stragglers $\alpha<200^\circ$ in our candidate list, most will be
classified blue straggler, and the reduction in velocity dispersion
would be greatest in our A/BS sample -- the opposite of what is seen.



\begin{figure}
\centering{
\scalebox{0.40}{
\includegraphics*{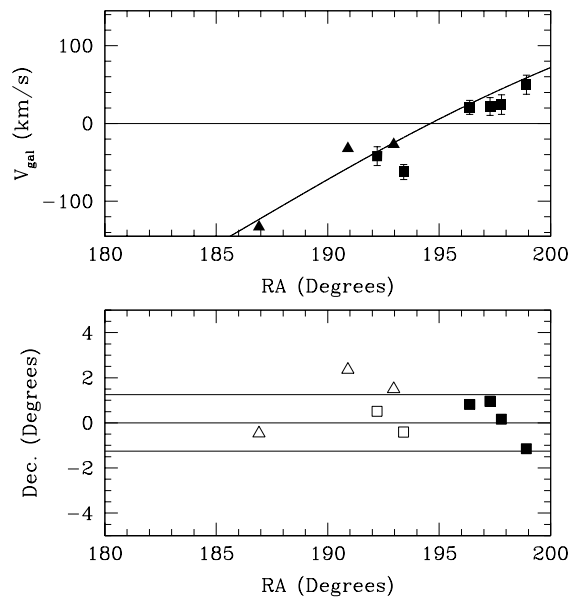}}
}
\caption{{\em Upper:} Plot of V$_{gal}$ versus RA for the six BHB
stars (shown as solid squares) and three carbon stars (solid
triangles) that are potential members of a stream. The plot suggests
that these stars are located near the turning point of an orbit (where
the radial velocity changes sign). The turning point is located at an
RA of about $195^\circ$. {\em Lower:} A plot of RA versus Dec. for the
stars in the stream. The filled symbols are the stars with V$_{gal}$
$>$ 0 and the open symbols are for V$_{gal}$ $<$ 0 otherwise the
symbols are the same as in the upper plot. The BHB stars are confined
to the northern stripe of the EDR data set shown by the horizontal lines.}
\label{turn_point}
\end{figure}

We conclude from the foregoing discussion that we have succeeded in
defining samples of BHB and A/BS stars, with small contamination, that
show distinct kinematic properties.  Indeed the fact that the measured
velocity dispersions of the two populations are significantly
different is confirmation of the reliability of the classification
methods.  In seeking an explanation for this difference, a number of
possible dynamical explanations could be pursued, for example that the
stellar orbits change from being predominantly radial to predominantly
circular at large radii (e.g. Sommer-Larsen et al.  1997). While this
is possible, a more convincing explanation was immediately apparent.

Six of the BHB stars are confined to a small region of space with
$190^\circ<\alpha<200^\circ$, $63<r<78\,$kpc (and a small range in
$\delta$).  The average distance of these stars from the centre of the
Galaxy is $70.6\,$kpc.  These stars are therefore confined to a very
small fraction of the volume surveyed.  The velocity dispersion of
these six stars (corrected for measurement errors),
$42\pm12$km\,s$^{-1}$, is too large to associate them with a bound
object i.e.  a low surface--brightness dwarf galaxy.  The key to
understanding the anomalous velocity dispersion is provided in
Figure~\ref{turn_point}(upper), which plots Galactocentric radial
velocity $V_{\rm gal}$ versus RA for these six BHB stars, marked by
squares.  A correlation is evident, indicative of streaming motion,
perhaps associated with a disrupted satellite.
Figure~\ref{turn_point}(lower) plots the position on the sky of the
six stars.  To investigate further the possibility of a stream, we
searched the catalogue of faint carbon stars of Totten \& Irwin (1998)
in the vicinity, confining ourselves to the ranges RA
$160^\circ<\alpha<200^\circ$, Dec.  $\pm$ $5^\circ$, and to similar
distances (improved distance estimates taken from Totten, Irwin \&
Whitelock, 2000).  Three stars meet these criteria, and have been
added to both diagrams in Figure~\ref{turn_point}, marked as
triangles.  Remarkably the three stars appear to add to the evidence
of a stream.  Details of the three stars are provided in Table
\ref{data_sum}.  (The velocity errors, taken from Totten, Irwin \&
Whitelock (2000), are 4, 4 and 6 km s$^{-1}$ in the order the stars
appear in the Table)

The correlation between $V_{\rm gal}$ and RA evident in
Figure~\ref{turn_point}(upper), encompasses $V_{\rm gal}=0$, which
would correspond to a turning point in the orbit at RA of $\sim
195^\circ$.  In order to investigate this trend in more detail, we
consider orbits in the spherical potential

\begin{equation}
\Psi(r) = \frac{GM}{a}\log\left[\frac{\sqrt{a^2+r^2}+a}{r} \right] .
\label{eq:pot}
\end{equation}

The scale length $a=178.0\,$kpc and the mass $M = 2.0\times 10^{12}$\,M$_\odot$
are chosen to match those estimated for the halo of the Milky Way (Wilkinson et
al.  2003).  We investigate whether the trend in Figure~\ref{turn_point} can be
reproduced by a plausible Galactic orbit as follows.  First, we assume that the
orbit has a turning point in the RA range $190-200^\circ$ and in the distance
range $50-80\,$kpc.  We then choose values for the line-of-sight distance $d_0$,
Galactic latitude $b_0$ and longitude $l_0$ of the turning point and the values
of the two components of velocity transverse to the line of sight, $v_{\rm b,0}$
and $v_{\rm l,0}$.  From each set of initial conditions, we integrate the orbit
in the RA range $190-200^\circ$ and determine $V_{\rm gal}(RA)$.  Assuming
Gaussian errors $\sigma_i$ on the individual radial velocities (and neglecting
any errors in the RA measurements), the probability that the data $(V_{\rm
gal,i},RA_i)$ were drawn from the relation $V_{\rm gal}(RA)$ is given by

\begin{eqnarray}
P(v_{\rm los,i},RA_i \vert l_0,b_0,r_0, v_{\rm b,0},v_{\rm l,0}) = 
\qquad\qquad\qquad\qquad&&\nonumber\\
\prod_i \frac{1}{\sqrt{2\pi\sigma_i^2}} \exp\left[-\frac{(v_{\rm
los}(RA_i) - v_{\rm los,i})^2}{2\sigma_i^2}\right].&&
\label{eq:orb_prob}
\end{eqnarray}
We use a downhill simplex algorithm (the routine \texttt{amoeba} in
Press et al. 1992) to maximise this probability over the five
dimensional parameter space. The resulting $V_{\rm los}(RA)$ relation
is shown in Figure~\ref{turn_point}. The orbit we obtain is strongly
radial and has an apocentre of $67.1\,$kpc and pericentre of
$7.5\,$kpc. The energy and angular momentum of the orbit are $E =
-8.1\times 10^4$(km\,s$^{-1})^2$ and $L^2 = 1.2\times 10^7\,$(kpc
km\,s$^{-1})^2$. The velocity dispersion of all nine stars relative to
this orbit is $15\pm 4\,$km s$^{-1}$.

Up to this point, we have not made use of the estimated distances to
our tracer stars, due to their relatively large uncertainties. We can
include them in the orbit determination in a straightforward manner by
multiplying the probability in eq~(\ref{eq:orb_prob}) by a second
Gaussian $P(d_{\rm los})$ given by
\begin{eqnarray}
P(d_{\rm los,i},RA_i \vert l_0, b_0,r_0, v_{\rm b,0},v_{\rm l,0}) =
\qquad\qquad\qquad\qquad &&\nonumber\\
\frac{1}{\sqrt{2\pi\sigma_{d,i}^2}} \exp\left[-\frac{(d_{\rm
los}(RA_i) - d_{\rm los,i})^2}{2\sigma_{d,i}^2}\right].&&
\end{eqnarray}

Here, $\sigma_{d,i}$ is the uncertainty in the line of sight distance to the
$i$th star.  In fact, the inclusion of this term results in an almost identical
orbit, the distance uncertainties rendering the distance estimates of little
value in the determination of orbital parameters.

We initially analysed the data using the technique of Lynden-Bell
\& Lynden-Bell (1995). These authors note that if a group of stars lie on the same orbit in an
assumed spherical potential, they share the same energy $E$ and total
angular momentum $L^2$. Since the energy is given by
\begin{equation}
\vert E\vert = \Psi(r) - \frac{1}{2} v_{\rm r}^2 - \frac{1}{2} \frac{L^2}{r^2},
\end{equation}
where $v_{\rm r}$ is the Galactocentric radial velocity, $L$ is the
magnitude of the total angular momentum vector and $\Psi(r)$ is the
potential, the stars in a stream lie on a straight line in a plot of
$\vert E_{\rm r}\vert$ versus $r^{-2}$ where
\begin{equation}
\vert E_{\rm r}\vert = \Psi(r) - \frac{1}{2} v_{\rm r}^2.
\end{equation}

If we apply this technique to our data, we obtain orbits with very significantly
higher orbital angular momenta than that for the orbit which fits the $V_{\rm
gal}$ versus RA relation above.  After consideration of the propagation of the
observational errors into the $(|E_r|,1/r^2)$ plane, we noted that the orbital
angular momentum was in fact being governed by the orientation of the extended
error ellipses caused by the large distance uncertainties.  In addition, the
derived orbit does not reproduce the trend of $V_{\rm gal}$ with RA seen in
Figure~\ref{turn_point}.  We conclude therefore that this technique can only
yield useful information about an orbit if the magnitudes of the distance errors
are significantly smaller than the radial range covered by the survey -- this is
not the case for our present sample.

Given the proximity of the Sagittarius (Sgr) stream, it is reasonable to suppose
the BHB stars might be part of a more distant passage of the stream around the
Milky Way (e.g.  Helmi \& White 2001; Dohm-Palmer et al.  2001).  Comparing the
positions and heliocentric velocities of our sample of stars to the simulations
plotted in Dohm-Palmer et al.  (2001; their figure 3) reveal that our sample
largely lies either between or beyond two wraps of the Sgr stream.  In fact, two
of our six BHB stream stars (numbers:  34 and 28) are inconsistent with the
model simulations.  More recently, Law, Johnston \& Majewski (2005) produced
models of the Sgr tidal tails using test particle orbits and $N$-body
simulations in a variety of potentials.  Before we are able to compare our data
with these models we need to convert our coordinates to the system defined in
Majewski et al.  (2003).  In this coordinate system the zero plane of the
latitude coordinate $B_{\odot}$ coincides with the best-fit great circle defined
by the Sgr debris, as seen from the Sun; the longitudinal coordinate
$\Lambda_{\odot}$ is zero in the direction of the Sgr core and increases along
the Sgr trailing stream.  Our sample resides in the region $249^\circ<
\Lambda_{\odot} <275^\circ$ and $7^\circ < B_{\odot} < 20^\circ$.  At these
coordinates the whole sample of stars in Table \ref{data_sum} resides at larger
distances than the models predict.  Clearly, the observation of a larger sample
of remote BHB stars in the Galaxy halo, along different lines of sight, is
essential to confirm the reality of the stream.  Additionally, we need to
establish whether the small velocity dispersion measured for the eight distant
BHB stars discovered in this paper is actually because six of the stars are
associated with a coherent structure, or because the velocity dispersion of the
whole population of outer halo BHB stars falls steeply with radius.  However, we
note that the apparent dominance of streaming motion in our BHB sample lends
support to the claim of Majewski (2004) that the non-uniform kinematics of outer
halo K-giants are consistent with that population having derived almost
completely from accretion.

We close with a summary of the main points of this paper. We have
presented the results of a survey of remote halo A-type stars selected
from the SDSS. Spectroscopy of the A-type stars obtained with the VLT
produced a sample of 20 stars with data of suitable quality for
classification into the classes BHB and A/BS. The final sample (Table
4) comprises eight stars classified BHB, at distances of $65-102\,$kpc
from the Sun (mean distance $80\,$kpc), with heliocentric radial
velocities accurate to 12 km s$^{-1}$, on average, and distance errors
$<10\%$. This is the most distant sample of Galactic stars with
measured radial velocities, of this size. Of the eight remote BHB
stars, we find that six show a strong trend in $V_{\rm gal}$ with RA,
and are consistent with a single orbit in a spherical halo
potential. The measured dispersion of the radial component of the
Galactocentric velocity for this sample is $42\pm12$km\,s$^{-1}$. This
value is significantly smaller than values measured for samples of
stars at smaller radii, and for satellites at similar radii.  This
evidence is supported by the existence of three previously identified
carbon stars with the same kinematics. A simple model shows all the
stars lying on an orbit with energy and angular momentum of $E =
-8.1\times 10^4$(km\,s$^{-1})^2$ and $L^2 = 1.2\times 10^7\,$(kpc
km\,s$^{-1})^2$. The velocity dispersion of the nine stars is
56$\pm$13km s$^{-1}$; the dispersion relative to the calculated orbit
is 15$\pm$4km s$^{-1}$. We conclude that we find a strong indication
of the presence of a stream but further observations are required to
trace the full extent of this stream on the sky.

\section*{Acknowledgements} 
We thank Kyle Cudworth for providing us with accurate positional
information for the radial velocity stars in M5, and J.A.Smith for
supplying us with photometric data. We also thank Mike Irwin for
several valuable discussions and for pointing us towards the carbon
star data set of Totten \& Irwin. The Sgr coordinate system
conversions made use of code at:
http://www.astro.virginia.edu/$^\sim$srm4n/Sgr/code.html. We are
grateful to the referee for comments that helped improved the clarity
of the manuscript.  LC and MIW acknowledge PPARC for financial
support. This paper uses observations made on the Very Large Telescope
at the European Southern Observatory, Cerro Paranal, Chile [programme
ID: 71.B-0124(A)]. We made use of the SDSS online database.  Funding
for the creation and distribution of the SDSS Archive has been
provided by the Alfred P. Sloan Foundation, the Participating
Institutions, the National Aeronautics and Space Administration, the
National Science Foundation, the U.S. Department of Energy, the
Japanese Monbukagakusho, and the Max Planck Society.

\end{document}